\pdfoutput=1
\documentclass[aps]{revtex4}
\usepackage[latin1]{inputenc}
\usepackage{graphicx,epsfig,rawfonts,pictex,multirow,longtable}

\def\etal{{\it et al.}}

\def\hlf{{{1\over2}}}
\def\thlf{{{3\over2}}}
\def\fhlf{{{5\over2}}}
\def\shlf{{{7\over2}}}
\def\>{\rangle}
\def\<{\langle}
\def\Schrodinger{Schr\"odinger }
\def\rmb#1{{\bf #1}}

\def\beq{\begin{equation}}
\def\eeq{\end{equation}}

\def\beqy{\begin{eqnarray}}
\def\eeqy{\end{eqnarray}}

\begin{document}

\pagenumbering{arabic}

\title
{Symmetries and Systematics of Doubly Heavy Hadrons}
\author{B. Eakins and W. Roberts}
\affiliation{Department of Physics, Florida State University, Tallahassee, FL 32306}
\begin{abstract}
We discuss the extension of the superflavor symmetry of doubly heavy baryons to states which contain an excited heavy diquark, and we examine some of the consequences of this symmetry for the spectra of doubly heavy baryons and heavy mesons.  We explore the ramifications of a proposed symmetry that relates heavy diquarks to doubly heavy mesons.  We present a method for determining how the excitation energy of a system containing two heavy quarks will scale as one changes the strength of the interactions and the reduced mass of the system.  We use this to derive consequences of the heavy diquark-doubly heavy meson symmetry.  We compare these consequences to the results of a quark model as well as the experimental data for doubly and singly heavy mesons.  We also discuss the possibility of treating the strange quark as a heavy quark and apply the ideas developed here to strange hadrons.
\end{abstract}

\maketitle

\section{Introduction and Motivation}
Hadrons containing two heavy quarks (hereinafter: doubly heavy) have the potential to provide deep insights into the details of confining forces.  Furthermore, these particles offer an opportunity to explore the symmetries which emerge in hadrons that contain more than one heavy quark.  An interesting possibility is that there is a symmetry that relates hadrons with two heavy quarks to hadrons containing a single heavy quark.  Savage and Wise were the first to propose this, and they used it to relate the hyperfine splittings of baryons containing two heavy quarks (doubly heavy baryons, or DHBs) to those of heavy mesons containing a single heavy quark \cite{SavageSpectrum}.  We refer to this as the superflavor symmetry.  Subsequently, this symmetry has been used to relate the hyperfine splittings of tetraquarks containing two heavy quarks to those of baryons containing a single heavy quark \cite{Cohen:2006jg}.

The fundamental notion of the superflavor symmetry is that, in the infinitely massive heavy quark limit, the two heavy quarks in a doubly heavy hadron will bind tightly into a heavy diquark in a color antitriplet, and the light degrees of freedom will  interact with this stationary, pointlike source of anticolor. As a result the details of nonperturbative QCD interactions involving the light degrees of freedom in a $QQq$ system will be identical to the corresponding interactions in a $\overline{Q}q$ system.  For the ground state, two heavy quarks of the same flavor will exist in a spin triplet due to Fermi statistics.  Savage and Wise went on to argue that the interactions between the two heavy quarks are dominated by a color Coulomb potential due to the small average separation between them.  Under this assumption, the energy of the heavy diquark scales as $\alpha_{s}^2(M_Q)M_Q$, in analogy to positronium.  This implies that, for sufficiently massive quarks, the energy of these excitations will be much larger than the energy required to excite the light degrees of freedom. However, several authors have obtained results for doubly heavy baryons which disagree with this estimate \cite{RobertsPervin, Fleck:1989mb, Stong, Gershtein:1998sx, gershtein00, Ebert:2002ig, Kiselev:Omegas, Majethiya, Giannuzzi}.  Indeed, the Roberts-Pervin (RP) model predicts that, as the heavy quarks become more massive, the energy required to excite the heavy diquark {\em decreases}.  If this is the case, then one cannot ignore these excitations, and it is desirable to extend the superflavor symmetry so that it incorporates this possibility.  This paper seeks to test the validity of such an extension in the context of a constituent quark model \cite{RobertsPervin}.  We also demonstrate the necessity of this extension of the superflavor symmetry by estimating the energy required to excite heavy diquarks using experimental data for doubly heavy mesons.

There are many articles that calculate DHB spectra.  Among these treatments are nonrelativistic constituent quark models (NRCQMs) which deal with the three body problem directly \cite{RobertsPervin, Stong, Rujula75, Fleck:1989mb, Bagan:1994dy, SilvestreBrac:1996wp, Richard:1996za, itoh00, vijande04, Albertus:2006ya, Zheng10}.  There are also several NRCQMs which solve the three-body problem through the Hyperspherical formalism  \cite{Majethiya, martynenko, Patel08, narodetskii02, Narodetskii:2002ks} and some which solve the problem through a quark-diquark approximation (we discuss this method in section \ref{factor}) \cite{Gershtein:1998sx, gershtein00, Kiselev:Omegas, Likhoded}.  There are relativistic quark models which treat DHBs, by a quark-diquark approximation \cite{Ebert:2002ig,Ebert:1996ec,Ebert:2005ip} and otherwise \cite{Gerasyuta:1999pc, Lyubovitskij:2003pn, faessler, Gerasyuta:2008zy}.  In addition, DHBs have been treated with bag models \cite{Fleck:1989mb, He2004}, the Bethe-Salpeter equation \cite{Giannuzzi, tong2000, Weng2011}, QCD sum rules \cite{Bagan:1994dy, Bagan:1992za, Kisilev2001, Zhang:2008, Alb:Nuc, Alb:PhLett, tang:2011}, effective field theories \cite{SavageSpectrum, Brambilla05, Hu2005, Fleming2006, Mehen2006}, Lattice QCD \cite{lewis01, Flynn:2003vz, Na:2007pv}, and by solving Non-Relativistic QCD on the lattice \cite{Mathur:2001id, mathur02}.  Predictions for the ground states of DHBs have also been made by exploiting regularities in hadron spectra and combining those regularities with inequalities derived through the Hellmann-Feynman Theorem \cite{Lichtenberg:1995kg, Roncaglia:1995az}.

We discuss six aspects of the spectra of doubly heavy hadrons. First, we develop an energy scaling rule for systems consisting of two heavy quarks, such as doubly heavy mesons and heavy diquarks,  which shows how the excitation energy scales with the reduced mass of the system.  We then explore a possible relationship between the excitation energies of heavy diquarks and those of doubly heavy mesons.  The third aspect is the expected existence of degenerate multiplets in the spectrum of DHBs, similar to those of Heavy Quark Symmetry.  The fourth aspect we discuss is the expectation that the energy required to excite the light degrees of freedom of a DHB will be independent of the specific excited state of the heavy diquark.  The fifth aspect is a heavy diquark flavor symmetry which relates the properties of $\Xi_{cc}$, $\Xi_{bc}$, and $\Xi_{bb}$ baryons.  The sixth aspect is a superflavor symmetry which relates DHBs with a heavy diquark in an arbitrary excited state to singly heavy mesons.  Finally, we examine how these six aspects emerge (or not) in baryons containing strange quarks, with the strange quark treated as a heavy quark.  We examine these ideas using results from a non-relativistic quark model \cite{RobertsPervin} and, when possible, experimental data.  Factorized (see section \ref{factor}), quark-diquark approaches to DHBs should naturally contain these DHB symmetries.  However, the model that we employ was not constructed with these symmetries and does not assume a quark-diquark structure.  Therefore, one of the goals of this manuscript is to investigate if these symmetries emerge in such a model.

\section{Heavy-Heavy Systems}

\subsection{Factorization} \label{factor}
We begin our discussion by exploring the physics of doubly heavy mesons and heavy diquarks. In order to accomplish this, we work in a limit where an approximation known as factorization \cite{kiselev02,gershtein00,Giannuzzi} is valid for DHBs.  If two quarks are sufficiently heavy the potential between them is expected to be essentially Coulombic, and the size of the $QQ$ subsystem of a DHB will decrease as $M_Q$ increases, with $r_{QQ} \sim 1/M_Q \ll 1/\Lambda_{QCD}$.  Factorization posits that, for a pointlike heavy diquark, one may first solve for the spectrum of the (heavy-heavy) diquark system, and then subsequently, solve for the energy of the heavy diquark-light quark (heavy-light) system.  Thus, a three body problem (in a constituent quark model picture) factorizes into two, independent two-body problems.  Unfortunately, the $c$ and $b$ quarks may not be massive enough for a diquark composed of these quarks to be considered a pointlike object bound by a color Coulomb potential \cite{Cohen:2006jg}. However, a pointlike diquark is not a necessary condition for factorization.  A simple example of this is a system of three particles confined by a harmonic oscillator potential.  The Hamiltonian factorizes explicitly after a suitable choice of coordinates, provided that the masses of the heavy quarks are equal \cite{Castro}.  This approximation complements our discussion about the symmetries of the heavy-light system in that factorization can easily incorporate the spin and (super)flavor symmetries that we will present in section \ref{Heavy-lightSymmetries}.  

Factorization may be viewed as a first approximation to a more careful treatment of the interactions between the light degrees of freedom and the heavy diquark.  In this approximation, corrections due to the finite size of the heavy diquark may be handled using form factors \cite{Ebert:2002ig}.  The validity of this approach also depends on whether confining forces are pairwise or three-body; factorization may be expected to be more accurate for three-body rather than pairwise confining forces \cite{kiselev02,Stong}.  For the discussion of factorization, we work in the limit where the ground state and the excited states of the heavy diquark are sufficiently pointlike so that we may neglect corrections due to its finite size.  

If factorization holds reasonably well for baryons containing $b$ and $c$ quarks, it may be possible to extract the spectra of heavy diquarks from experimental data.  This may make it possible to investigate the symmetries of a bound state of two heavy quarks in an {\em antitriplet} color configuration, with the understanding that this color charge is screened by the light degrees of freedom.  Howbeit, these screening effects do not significantly alter the physics of the heavy diquark system if the diquark is sufficiently pointlike.  Thus, the interactions between the quarks in the $QQ$ system of a DHB should be similar to those between the quark and antiquark of a doubly heavy meson \cite{gershtein00}.  The primary difference arises from the color factors.  Na\"ively, these factors are the same for every type of interaction involved, implying that the strength of interactions in the doubly heavy meson system are twice that of interactions in the heavy diquark system.  This is true at tree level in perturbative QCD, and there are indications from quenched latice QCD that this is also true of the long-distance part of the confining potential \cite{Nakamura}.  If this holds rigorously, then it amounts to a symmetry between color antitriplet and color singlet bound states containing two heavy quarks.

\subsection{Interaction Strength and Mass Scaling}
The appropriate effective theory for exploring heavy-heavy systems is nonrelativistic QCD (NRQCD) \cite{Brambilla05}, but here we choose a simpler and more familiar approach.  We work in the limit where the two heavy quarks are massive enough that their motion can be treated as essentially non-relativistic and QCD interactions can be well-described by an adiabatic potential.  In other words, we are in the limit where the \Schrodinger equation is valid.  For simplicity, we also ignore all spin-dependent interactions and assume a flavor independent potential.  The most general Hamiltonian for the excitation energy of two heavy quarks that can be written for this situation is 
\beq \label{SimpleKindOfHam}
H=\frac{P^2}{2\mu}+bV(r),
\eeq
\noindent where $r$ is the relative coordinate between the two heavy quarks, $\mu$ is the reduced mass of the system, and $P$ is the momentum conjugate to $r$.  We have also included a strength parameter, $b$, as an explicit factor multiplying the potential, $V(r)$.  This Hamiltonian does not mix states with different values of orbital angular momentum because there are no scalar spin-independent interactions which can do so.  Our goal is not to solve the eigenenergy problem explicitly, but to discern how the eigenenergies scale as one changes the reduced mass and the strength parameter.  The appropriate tool for accomplishing this task is the Hellmann-Feynman Theorem, which has been applied previously to hadron systems \cite{Bagan:1994dy, Lichtenberg:1995kg, Roncaglia:1995az, Quigg, Cohen1979, Kwong, Lipkin1993, Roncaglia}, but our approach is somewhat novel.

The Hellmann-Feynman theorem states that, if a Hamiltonian, H, its $nth$ eigenenergy, $E_n$, and the $nth$ eigenstate, $| \Psi_n \>$, are differentiable functions of a parameter, $\lambda$, then the energy satisfies,
\beq
\frac{\partial E_{n}}{\partial \lambda}=\< \Psi_n |\frac{\partial H}{\partial \lambda} | \Psi_n \>.
\eeq
\noindent For simplicity, we drop all of the labels referring to the eigenstate and remove all explicit references to the eigenfunctions.  An application of this theorem to the Hamiltonian of eqn. (\ref{SimpleKindOfHam}) leads to
\beq
\frac{\partial E}{\partial \mu}=-\frac{1}{2 \mu^2}\< P^2 \>, \text{ and}
\eeq
\beq
\frac{\partial E}{\partial b}=\< V(r) \>.
\eeq
\noindent On the other hand, the eigenenergy may be written as,
\beq
E=\frac{\<P^2\>}{2\mu}+b\<V(r)\>.
\eeq
\noindent After a simple substitution, we obtain
\beq
E=-\mu \frac{\partial E}{\partial \mu} + b\frac{\partial E}{\partial b}.
\eeq
\noindent Provided that $E \ne 0$, this partial differential equation is separable.  In this case, the solution to the PDE is
\beq \label{Escale}
E_{nl}=D_{nl} \mu^{d-1}b^d,
\eeq
\noindent where $D_{nl}$ is independent of $\mu$ and $b$, $n$ labels the radial excitation, $l$ labels the orbital excitation and  $d$ is a constant index to be determined, which depends on the specific potential that is used, but is independent  of the particular eigenstate, labeled by $n$ and $l$. We show that this is true for a simple case below.  This equation was derived previously by Quigg and Rosner \cite{Quigg} for the special case of a power law potential.  For a potential which is positive definite, the solution is exact. However, if the potential, $V(r_0)=0$ for some $r_0 \in (0,\infty)$, then there may be some value of the reduced mass for which the eigenenergy crosses zero.  In this respect the solution above can be regarded as an approximation where one chooses the value of $d$ so that the equation reproduces the spectrum for a limited range of reduced masses.  It may be possible to find a more general solution to the PDE above, but this form of the solution, eqn. (\ref{Escale}), is quite convenient for our purposes.

\subsection{Scaling for Special Cases}
To illustrate the use of the energy scaling rule (eqn. \ref{Escale}), consider an attractive power law potential,
\beq
V(r)=\text{sgn}(\nu)r^\nu.
\eeq
\noindent An application of the Quantum Virial Theorem yields,
\beq
\<P^2\>=b\mu \nu \<V(r)\>.
\eeq
\noindent This result can be combined with the results from the Hellmann-Feynman Theorem to obtain,
\beq
E=\frac{2+\nu}{2} b \frac{\partial E}{\partial b}.
\eeq
\noindent The solution to this equation scales as $b^{\frac{2}{2+\nu}}$, and so
\beq
d=\frac{2}{2+\nu}
\eeq
\noindent for a power law potential.  For the special case of a Coulomb potential, the energy is proportional to the reduced mass, while the energy for a linear potential scales as $\mu^{-{1\over3}}$, and the energy for a harmonic oscillator scales as $\mu^{-{1\over2}}$.  This simple example yields valuable intuition about the meaning of the scaling index, $d$; if $d \approx {\frac{2}{3}}$ for a range of reduced masses then the potential is dominated by a linear term, but if $d \approx 2$ then the potential is effectively Coulombic.  This special case for the energy scaling equation of a power law potential agrees with the result which Quigg and Rosner obtained using another method \cite{Quigg}.
\subsection{Doubly Heavy Meson Ratios}
The energy scaling rule, eqn. (\ref{Escale}), provides a simple way to analyze how orbital and radial excitations of a heavy-heavy system scale with the mass of the heavy quarks.  Because $D_{nl}$ is independent of the reduced mass, $\mu$, the scaling equation predicts that 
\beq \label{DHMratio}
\frac{M^{n'l'}_{Q_{1}\overline{Q}_{2}}-M^{nl}_{Q_{1}\overline{Q}_{2}}}  
     {M^{n'l'}_{Q'_{1}\overline{Q}'_{2}}-M^{nl}_{Q'_{1}\overline{Q}'_{2}}}
      = \left( \frac{\mu_{Q_1 Q_2}}{\mu_{Q'_1 Q'_2}} \right)^{d-1},
\eeq
\noindent where $M^{nl}_{Q_{1}\overline{Q}_{2}}$ is the mass of a doubly heavy meson with radial and orbital quantum numbers $n$ and $l$, respectively, and quark flavors $Q_{1}$ and $Q_{2}$.  $\mu_{Q_1 Q_2}$ is the reduced mass of the $Q_1 \overline{Q}_2$ system.  This expression is only reliable in the nonrelativistic, adiabatic limit, and it contains an inherent ambiguity, in that the reduced mass of the heavy quark system cannot be directly measured by experiment.  The values of the reduced masses will be model dependent, and models which allow for running masses should evaluate each reduced mass at the appropriate renormalization scale.  Corrections due to the running of the coupling constant may also be warranted.  These ratios also ignore spin-dependent interactions and, as such, they should be approximately valid for spin-averaged masses.  The striking prediction that this relation makes is that a host of ratios involving {\em different} excited states of doubly heavy mesons will be identical:  no matter what choices are made for $n$, $n'$, $l$, and $l'$ the ratio is the same.  If radial and orbital splittings for doubly heavy mesons satisfy this prediction, then it will support the reasonableness of the energy scaling rule, eqn. (\ref{Escale}).

\section{Heavy-light Systems} \label{Heavy-lightSymmetries}

\subsection{Total Angular Momentum Decoupling}
In this discussion we use the successful Heavy Quark Symmetry (HQS) \cite{IsgurWise} as a guide for exploring the symmetries which might emerge for doubly heavy hadrons.  For hadrons containing a single, infinitely massive heavy quark, the spin of the heavy quark decouples from the spin of the light degrees of freedom.  Thus, singly heavy hadrons exist in degenerate doublets with total angular momenta $J=J_l \pm s_Q$, where $s_Q=\frac{1}{2}$ is the spin of the heavy quark and $J_l$ is the total angular momentum of the light degrees of freedom.  For doubly heavy hadrons, it is a natural extension of this concept to expect that the total angular momentum of the heavy diquark should decouple from the total angular momentum of the light degrees of freedom leading to degenerate multiplets with total angular momenta $J$ satisfying
\beq
|J_d-J_l| \le J \le J_d + J_l,
\eeq
\noindent where $J_d$ is the total angular momentum of the heavy diquark.  If this relation is only applied to situations where $J^{\pi_d}_d=1^+$ or $0^+$, then one recovers the symmetry proposed by Savage and Wise \cite{SavageSpectrum}. However, their formalism may easily be extended to diquarks with higher spin \cite{GeorgiFlav}.

\subsection{Flavor and Superflavor Symmetries} \label{Superflav}
For singly heavy hadrons the excitation spectrum of the light degrees of freedom becomes independent of the heavy quark mass, in the infinitely massive heavy quark limit.  Thus, the excitation spectrum of a hadron containing a $c$ quark is very similar to the excitation spectrum of a hadron containing a $b$ quark. For doubly heavy hadrons, however, this symmetry is broken because a consistent effective field theory requires the inclusion of kinetic energy terms for the $b$ and $c$ quarks which depend explicitly on their masses \cite{Thacker91}.  The symmetry that remains for doubly heavy systems is the superflavor symmetry \cite{SavageSpectrum}.  If the heavy diquark is indeed a pointlike, stationary source of anticolor, one may construct a new effective theory where the relevant degrees of freedom are the heavy diquark and the light degrees of freedom. In this theory one may neglect the kinetic energy of the heavy diquark.  More specifically, interactions which change the velocity of the heavy diquark are suppressed by its mass, $M_d \gg \Lambda_{QCD}$.

In analogy with the heavy quark effective theory (HQET), it is expected that the effective Lagrangian for the quark-diquark system will be independent of the diquark's mass. Thus, if a diquark of mass $M_d$ and total angular momentum and parity $J^{\pi_d}_d$ is replaced with a diquark of mass $M'_d$ and total angular momentum and parity $J'^{\pi'_d}_d$, the physics of the light degrees of freedom will be unchanged.  Additionally, the fact that the effective Lagrangian does not depend on the heavy diquark's spin (also in analogy with HQET) indicates that one may also replace a heavy diquark (of any angular momentum and parity) with a heavy antiquark, and the physics of the heavy-light system will be unaffected.  In practice this is a superflavor symmetry which relates the spectra and properties of singly heavy mesons to those of DHBs, independent of the excitation state of the diquark.  This is the generalization of the superflavor symmetry to which we alluded earlier.

Given the arguments above, a new perspective on HQS appears:  Heavy Quark Symmetry may be a special case of a larger symmetry between hadrons containing $N_Q$ heavy quarks and/or antiquarks in a color configuration $C$, interacting with $N_l$ light quarks in a color configuration $\overline{C}$, and hadrons containing $N'_Q$ heavy quarks and/or antiquarks in a color configuration $C$, interacting with $N_l$ light quarks in a color configuration $\overline{C}$.  In short, we speculate that the decoupling of the spin and flavor/mass of the heavy degrees of freedom from those of the light degrees of freedom may be a general feature of QCD in the infinitely massive heavy quark limit.  The validity of this generalization of HQS can be tested with heavy exotic hadrons, if they exist \cite{Cohen:2006jg}.

\section{Symmetry Constraints}
\subsection{Meson-Baryon Ratios}
If there is a symmetry between color singlet and color antitriplet bound states, then the spectra of DHBs and the spectra of doubly heavy mesons can be related in a simple way.  Factorization of DHBs implies that one may write the mass as the sum of the energy contributions from the heavy diquark and the light degrees of freedom,  $M=E_d+E_l$, and after combining this relation with the energy scaling rule eqn. (\ref{Escale}), we find that
\beq \label{MBratio}
\frac{M^{n'l'}_{Q_{1}\overline{Q}_{2}}-M^{nl}_{Q_{1}\overline{Q}_{2}}}  
     {M^{n' l'\nu\lambda}_{Q_{1} Q_{2} q}-M^{n l\nu\lambda}_{Q_{1}Q_{2} q} } = 2^d,
\eeq
\noindent where $M^{nl}_{Q_{1}\overline{Q}_{2}}$ is the mass of a doubly heavy meson with radial and orbital quantum numbers $n$ and $l$, respectively, and $M^{n l\nu\lambda}_{Q_{1} Q_{2} q}$ is the mass of a DHB which contains a heavy diquark in a state labeled by $n$ and $l$, and light degrees of freedom in a state labeled by $\nu$ and $\lambda$.  The eigenstate of the light degrees of freedom for the ``primed" DHB must be the same as that of the ``un-primed" DHB in order to cancel its contribution to their masses in the denominator of eqn. (\ref{MBratio}). This result assumes that the strength of the interaction between the heavy quark and heavy antiquark in the doubly heavy meson is exactly twice the strength of the interaction between the two heavy quarks in the DHB (Sec. \ref{factor}).  We have neglected the running of the coupling constant as well, but interactions in the heavy diquark system and interactions in the doubly heavy meson system should occur at a similar renormalization scale.  The expression also neglects spin interactions.  We also remark that, inasmuch as factorization is valid, this relation may be applied to $\Omega_{QQ'}$ as well as $\Xi_{QQ'}$ baryons.  Because this constraint allows one to compare properties of a meson to a baryon through their color structures, the combination of factorization with the symmetry between singlet and antitriplet color configurations might be thought of as a ``supercolor'' symmetry.

\subsection{DHB Spectrum}
The extension of the superflavor symmetry from section \ref{Superflav} stipulates that the physics of the light degrees of freedom for a DHB is independent of the flavor and excited state of the heavy diquark.  This means that splittings between the ground state and an excited state of a singly heavy meson should be the same as the analogous splitting in the light degrees of freedom of a DHB.  The DHB spectrum should consist of a pattern of states that repeats for each heavy diquark with total angular momentum and parity $J_d^{\pi_d}$.  Furthermore, this repeating pattern will be nearly identical to the singly heavy meson spectrum, after corrections due to spin interactions and the finite heavy quark mass have been made.  The decoupling of the total angular momentum of the heavy diquark will also lead to a DHB spectrum consisting of degenerate multiplets with total angular momentum, $|J_d-J_l|,...,J_d+J_l$.
\begin{center}
\begin{figure}
\caption{DHB Symmetries.  A DHB with two heavy quarks, $Q$ and a light quark $q$ can be treated as two independent systems due to factorization and the decoupling of the total angular momentum of the heavy diquark $d$.  The spectrum of the heavy-light $d-q$ system can be related to the spectrum of heavy mesons through the superflavor symmetry.  Additionally, the $d-q$ system can be related to a $d'-q$ (where $d'$ is a heavy diquark of a different flavor) system through the diquark flavor symmetry.  The spectrum of the heavy-heavy $Q-Q$ system can be related to the spectrum of doubly heavy mesons through the supercolor symmetry. 
\label{DHBfig}}
\includegraphics[width=1.0\textwidth,height=0.3733\textheight]{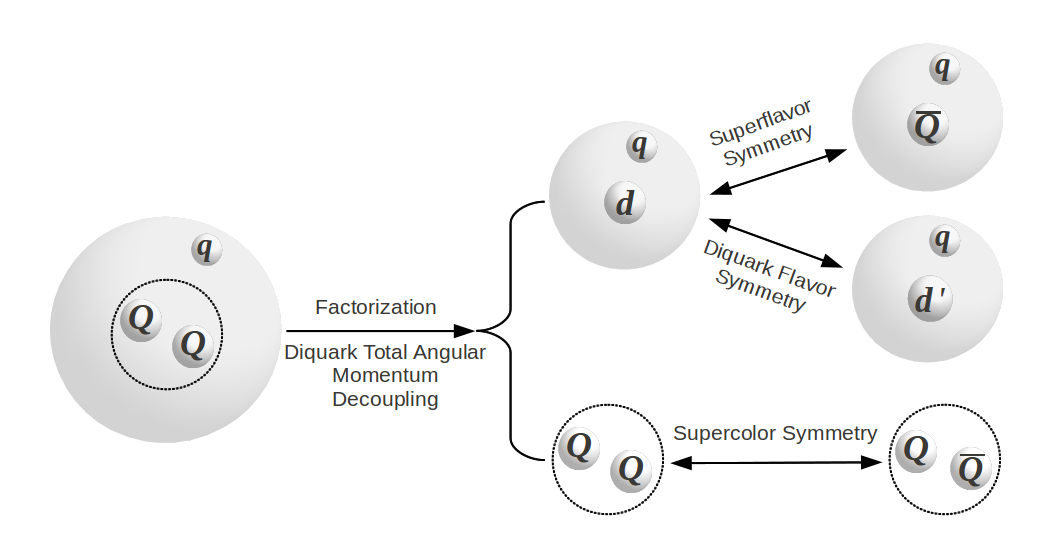}
\end{figure}
\end{center}

\section{Quark Model for DHBs} \label{QuarkModel}
\subsection{Hamiltonian}
In this section we present a non-relativistic quark model which Roberts and Pervin \cite{RobertsPervin} used to calculate masses of heavy baryons.  The advantage of this model is that it is among only a handful of articles \cite{RobertsPervin, Ebert:2002ig, gershtein00, Majethiya} in which orbital excitations both in the heavy diquark and in the light degrees of freedom are allowed.  Among these models, it is the only one which solves the three-body problem without using a quark-diquark approximation.  The model is also convenient for this manuscript because we have access to the wave functions and can comment on the multiplet structure of the states obtained in the spectrum.  For the purposes of this note, the spectrum of \cite{RobertsPervin} has been recalculated with spin-dependent interactions removed. The Hamiltonian is
\beq
H= K +V_{\rm conf}+ C_{qqq}.
\eeq
\noindent $C_{qqq}$ is a constant term which is the same for all baryons in the model.  The kinetic energy term is 
\beq
K=\sum_{i=1,3}\left( m_i+\frac{p_i^2}{2m_i} \right),
\eeq
\noindent and it may be factorized easily through the introduction of Jacobi coordinates,
\beq
{\bf \rho}= \frac{1}{\sqrt2}({\bf r}_1 - {\bf r}_2)
\eeq
and 
\beq
{\bf \lambda} =\sqrt{\frac{2}{3}}\left(\frac{m_1{\bf r}_1 + m_2{\bf r}_2}{m_1+m_2} - 
{\bf r}_3\right).
\eeq
\noindent With this choice, $\rho$ is proportional to the relative coordinate between the two heavy quarks, and $\lambda$ is proportional to the relative coordinate between the center of mass of the heavy diquark and the the light quark. In the rest frame of the baryon, the kinetic energy operator becomes
\beq
K= \sum_{i=1,3}m_i+\frac{p_\rho^2}{2\mu_{\rho}} + \frac{p_\lambda^2}{2\mu_{\lambda}},
\eeq
\noindent with $\mu_\rho=\frac{2 m_1 m_2}{m_1+m_2}$ and $\mu_\lambda=\frac{3}{2}\frac{(m_1+m_2)m_3}{(m_1+m_2)+m_3}$.  This particular model contains pairwise interactions for the confining potential,
\beq
V_{\rm conf}= \sum_{i<j=1}^3\left({\alpha_{\rm Lin}r_{ij}\over 2}-
{2\alpha_{\rm Coul}\over3r_{ij}}\right),
\eeq
\noindent with $r_{ij}=\vert\rmb{r}_i-\rmb{r}_j \vert$.  The $r_{12}$ terms combined with the $\rho$ term in the kinetic energy give a Hamiltonian for the heavy-heavy subsystem which is in the form of eqn. (\ref{SimpleKindOfHam}), the form we used earlier to derive the energy scaling rule, eqn. (\ref{Escale}).  Unfortunately, the terms in $r_{13}$ and $r_{23}$ depend on both $\rho$ and $\lambda$.  The model therefore does not factorize explicitly, nor is there any obvious limit in which factorization emerges.

%
%

Obtaining a good fit to the baryon spectrum for this model required setting the strength of the Coulomb parameter, $\alpha_{\rm Coul}$, to be quite small ($\sim 10^{-4}$).  As such, we expect the eigenenergy contribution from the ``$\rho$" part of the Hamiltonian to scale roughly as $\mu_\rho^{-{1\over3}}$, which explains why the model predicts that the energy required to excite the heavy diquark decreases as the heavy quark mass increases.  The small value of $\alpha_{\rm Coul}$ also has implications for the size of heavy diquarks in this model.  For a Coulomb-plus-linear potential one expects that the size of the heavy diquark approaches $r_{QQ} \sim 1/M_Q$ as the quarks become very heavy since the Coulomb term dominates. Similarly, the linear term dominates as the two quarks that comprise the diquark become very light, and the diquark size approaches $r_{QQ} \sim 1/(M_Q\alpha_{\rm Lin})^{1\over3}$.  Because the potential in this model is essentially linear, the RP model might be considered a worst-case scenario for factorization because it contains pairwise forces and because the heavy diquarks are less pointlike than those of other Coulomb-plus-linear NRCQMs.  We have listed the parameters of the model in table \ref{parameter1}.
\begin{center}
\begin{table}[h]
\caption{Parameters of the RP model \cite{RobertsPervin}. $m_\sigma$ is the mass of the $u$ and $d$ quark.  \label{parameter1}}
\vspace{2mm}
\begin{tabular}{ccccccc}
\hline \toprule
 $m_\sigma$ & $m_s$  & $m_c$  & $m_b$  & $\alpha_{\rm Lin}$& $\alpha_{\rm Coul}$ & $C_{qqq}$  \\
 (GeV)&(GeV)&(GeV)&(GeV)&(GeV$^2$)&  --  &(GeV)\\ \hline
0.2848& 0.5553&  1.8182 & 5.2019 &  0.1540 &$\approx 0.0$&-1.4204\\
\toprule
\end{tabular}
\end{table}
\end{center}

\subsection{Flavor Multiplets}\label{flavormultiplets}

Assuming that only the $b$ and $c$ quarks can be treated as heavy, states containing two heavy quarks can be placed in SU(2) multiplets. These multiplets are a triplet
\begin{equation}
 \Xi_{cc}=ccq,\,\,\,\,\,\Xi_{bb}=bbq,\,\,\,\,\Xi_{bc}=\frac{1}{\sqrt{2}}\left(cb+bc\right)q,
\end{equation}
and a singlet
\begin{equation}
\Xi^\prime_{bc}=\frac{1}{\sqrt{2}}\left(cb-bc\right)q,
\end{equation}
with the light quark $q$ being one of $u$, $d$ or $s$. Of course, this is only meant to be a classification scheme, but it has implications for the angular momentum structure of the states, since the spin-space-flavor wave function must be symmetric in `identical' quarks. For instance, if there is no excitation of the heavy diquark, the diquarks in the triplet must have spin 1, while those in the singlet must have spin 0.  

\subsection{Angular Momentum Supermultiplets}\label{Multiplets}
In the quark model presented above, states are defined in what we may call an $L$--$S$ basis.  Symbolically, the angular momenta in the states are coupled in the scheme $\{[l_\rho l_\lambda]_L [s_d s_l]_S\}_J$, where $l_\rho$ is the relative orbital angular momentum between the two heavy quarks, $l_\lambda$ is the relative orbital angular momentum between the light quark and the heavy diquark, $s_d$ is the spin of the heavy diquark, and $s_l=\frac{1}{2}$ is the spin of the light quark.  For the symmetry scheme discussed in Section \ref{Heavy-lightSymmetries}, states are identified by the spin and parity of the diquark, $J_{d}^{\pi_d}$, the spin and parity of the light degrees of freedom, $J_{l}^{\pi_l}$, and the total angular momentum and parity of the baryon, $J^P$.  In order to compare these states with quark model states it is necessary to transform the states from the $L$--$S$ basis to the $J_d$--$J_l$ basis, in which the angular momenta are coupled symbolically as $\{[l_\rho s_d]_{J_d} [l_\lambda s_l]_{J_l}\}_J$.  These two bases are related by a 9--j symbol,
\beq\label{HQSSstates}
 \sqrt{(2L+1)(2S+1)(2J_d+1)(2J_l+1)}
\left\{ \begin{array} {ccc} l_\rho & l_\lambda   & L \\ 
                            s_d & \frac{1}{2} & S \\
                            J_d & J_l   & J \end{array} \right\}.
\eeq
\noindent Using this, we can assign states obtained in the quark model to angular momentum supermultiplets by examining their wave function components.  
\subsection{Pair-wise Mixing}\label{pairmix}
The extension of the superflavor symmetry that we have proposed is broken in this model even when spin-dependent interactions are absent.  This arises because of pair-wise forces and the composite nature of the heavy diquark, and it is a feature of any model that includes such forces.  To illustrate, let us consider a scalar, pair-wise interaction $f(|\vec{r}_{ij}|)$ between quarks $i$ and $j$.  Such an interaction may be written quite generally as,
\begin{eqnarray} \label{pairwise}
f(|\vec{r}_{ij}|)=\sum^{\infty}_{l=0} g_l(\rho,\lambda) (Y_l(\Omega_\rho) \cdot Y_l(\Omega_\lambda)).
\end{eqnarray}
\noindent This expression comes from writing $\vec{r}_{13}$ and $\vec{r}_{23}$ as linear combinations of $\vec{\rho}$ and $\vec{\lambda}$ and expanding the function $f$ as a sum over spherical harmonics multiplied by a scalar function $g_l$.  For a Coulombic interaction, this equation yields the familiar multipole expansion. Matrix elements of the interaction $f(|\vec{r}_{ij}|)$ are proportional to a $6-j$ symbol,
\begin{eqnarray}
\label{Coulin6j}
\<l'_\rho l'_\lambda ; L| f(|\vec{r}_i-\vec{r}_j|)|l_\rho l_\lambda ; L\> \propto
\left\{ \begin{array} {ccc} l'_\rho & l_\rho   & l \\ 
                            l_\lambda & l'_\lambda & L  \end{array} \right\}.
\end{eqnarray}
\noindent  This operator, eqn. (\ref{pairwise}), is a scalar in the total orbital angular momentum of the system, $L$. It therefore mixes states having different values of $l_\rho$ and $l_\lambda$, but the same total $L$.  For instance, states having $l_\rho=0$, $l_\lambda=2$ mix with states having $l_\rho=2$, $l_\lambda=0$.   However, even if we ignore mixing between states with different values of $l_\rho$ and $l_\lambda$, the interaction is diagonal in the L--S basis, implying that it mixes states in the $J_d$--$J_l$ basis.  In the absence of other interactions, the mixing coefficients are independent of the mass of the heavy quarks and the strength of the interaction, and are given by eqn. (\ref{HQSSstates}). In general, this mixing also leads to a mass splitting between states that occupy the same symmetry multiplet, and thereby lifts the degeneracy of states within the multiplet.  However, if $l_\rho=l'_\rho=0$ {\em or} $l_\lambda=l'_\lambda=0$, the matrix elements of pairwise operators simplify because the 6--J symbol of eqn. (\ref{Coulin6j}) dictates that $l=0$.  The matrix element of the interaction is then diagonal in the $J_d$--$J_l$ basis, as well as the $L$--$S$ basis. The upshot is that these pairwise interactions only remove the degeneracies within multiplets which contain both an orbitally excited heavy diquark {\em and} orbitally excited light degrees of freedom.

\section{Results} \label{results}

It must be emphasized that the majority of treatments of DHBs assume factorization, so that most of the features that we have discussed arise naturally. For instance, the fact that the excitation energies of the light quark are independent of which state the heavy diquark is in, is a natural consequence of factorization. So, too, is the independence of these excitation energies of the flavor of the diquark. The angular momentum `supermultiplets' should also arise naturally in factorized models. Our goal is to elucidate whether or not models that do not assume factorization, or, to be more precise, how a particular model that does not assume factorization, manifests these predictions.

\subsection{Meson-Meson Ratios}

\label{mesonmeson}

The meson-meson ratios of eqn. (\ref{DHMratio}) provide an opportunity to determine the energy scaling index $d$ using experimental results.  There are almost enough experimental data to compute the spin averaged masses of the ground state, the first orbital excitation, and the first radial excitation for charmonium and bottomonium \cite{PDG}.  For the $h_b(1P)$, we use the recent result from the Belle Collaboration \cite{Belle}: for the ground state bottomonium, we use the measurement from the BaBar Collaboration \cite{BABAR}; and for the $\eta_b(2S)$, we use the result from the model of Ebert {\it et al.} \cite{EbertMesons}.  With these data there are two ratios that can be constructed: the ratio of the orbital splitting of charmonium to that of bottomonium, and a corresponding ratio for radial splittings.  The results are shown in table \ref{Meson-MesonTable}.  The energy scaling rule, eqn. (\ref{Escale}), predicts that these two ratios should be identical, and they are up to $6\%$.  This suggests that the assumptions behind the derivation of this scaling relation (non-relativistic dynamics, constant scaling index for a finite range of reduced masses), are reasonable.  The experimental ratios also encapsulate a non-trivial fact about the spectra of doubly heavy mesons --  the cost of an orbital or a radial excitation is nearly independent of the mass of the two heavy quarks.  This fact forces non-relativistic models to choose a potential such that the energy scaling parameter, $d \approx 1$ for the reduced masses of charmonium and bottomonium. This is quite different from the energy scaling for a Coulomb potential, and contradicts some formulations of NRQCD which use a power counting scheme that assumes Coulombic excitation energies \cite{BrambillaRev}.  Quigg and Rosner have shown previously that the energy splittings of doubly heavy mesons can be reproduced using a suitably tuned Coulomb-plus-linear potential or a logarithmic potential \cite{Quigg}.
\begin{center} 
\begin{table}[h]
\caption{Spin-averaged mass splittings for bottomonium and charmonium from experiment \cite{PDG}.  The energy scaling rule, eqn. (\ref{Escale}), predicts that the ratios of these splittings should be the same value, $\left(\frac{m_c}{m_b}\right)^{d-1}$.  This allows us to extract the energy scaling index, $d$, from these ratios using the heavy quark masses from the RP model \cite{RobertsPervin}.
\label{Meson-MesonTable}}
\vspace{2mm}
\begin{tabular}{ccccc}
\hline \toprule
Splitting \hspace{3pt}&\hspace{3pt}$\Delta m_{c \overline{c}}$ (MeV)\hspace{3pt}&\hspace{3pt}$\Delta m_{b \overline{b}}$ (MeV)\hspace{3pt}&\hspace{3pt}$\Delta m_{c \overline{c}}/\Delta m_{b \overline{b}}$ \hspace{3pt}&   $d$ \\ \hline
$M(1P)-M(1S)$  &    457.6 $\pm$ 0.31     &  457.0 $\pm$ 1.5      &          1.001 $\pm$ 0.003                &  0.999 $\pm$ 0.007 \\ 
$M(2S)-M(1S)$  &    606.1 $\pm$ 1.0      &  573.3 $\pm$ 1.6      &          1.057 $\pm$ 0.003                &  0.947 $\pm$ 0.007 \\ 
\toprule
\end{tabular}
\end{table}
\end{center}
\subsection{Meson-Baryon Ratios} \label{secMBrat}

The meson-baryon ratios from eqn. (\ref{MBratio}) provide a means of comparing DHBs and doubly heavy mesons, and these ratios are sensitive to the form of the confining potential.  We use this ratio to compare two different ``models''.  We construct ratios involving DHB masses from the RP model \cite{RobertsPervin} and the results of a simple model for doubly heavy mesons using the Hamiltonian, eqn. (\ref{SimpleKindOfHam}), with its parameters and potential set to match the RP model (see table \ref{Meson-BaryonTable}).  The RP model violates the meson-baryon ratios, eqn. (\ref{MBratio}), for a linear potential by 23\% or less.  However, their potential is not purely linear in the heavy diquark's relative coordinate, and the values for the ratios in the table above are all near 1.26.  This suggests that effects due to the three-body nature of the DHB system have modified the effective interaction strength of the QQ subsystem by a factor of 1.41.  This may be due to screening effects from the light quark, or it may be a geometrical factor arising because of the pairwise confining potential used in the model.

The meson-baryon ratios also provide a convenient way to estimate the orbital and radial splittings of heavy diquarks.  Using $d=1$ for the scaling index, we find that the radial splitting for heavy diquarks should be roughly 300 MeV and the orbital splitting should be about 230 MeV.  Because the energy scaling index $d \approx 1$ these estimates do not depend on quark model masses, and in this sense they are model independent.  These estimates assume that screening effects from the light degrees of freedom are negligible.    For the RP model the radial splittings of the $\Xi_{cc}$, $\Xi_{bc}$, and $\Xi_{bb}$, diquarks are 332, 294, and 223 MeV, respectively;  likewise, the orbital splittings are 188, 167, and 129 MeV.  This disagreement with the doubly heavy meson spectra is due to the $\mu_{d}^{-\frac{1}{3}}$ scaling of excitation energies in the RP model. 
\begin{center} 
\begin{table}[h]
\caption{Meson-Baryon ratios (eqn. \ref{MBratio}) from the spin independent RP model and a simple quark model for doubly heavy mesons.  For all of these ratios the light degrees of freedom are in the ground state.  The subscript $m$ refers to a meson and the subscript, $b$, refers to a baryon.  The ratios of these splittings should be the same value, $2^{2 \over3}=1.59$.
\label{Meson-BaryonTable}}
\vspace{2mm}
\renewcommand{\arraystretch}{2.1}
\begin{tabular}{cccc}
\hline \toprule
                        Ratio ($n_d l_d$)                   &  $cc$      & $bc$      & $bb$   \\ \hline
${\displaystyle \frac{M_m(1 P)-M_m(1 S)}{M_b(1 P)-M_b(1 S)}}$   &  1.28      &  1.26     &  1.32   \\
${\displaystyle \frac{M_m(2 S)-M_m(1 S)}{M_b(2 S)-M_b(1 S)}}$   &  1.24      &  1.23     &  1.30   \\
${\displaystyle \frac{M_m(1 D)-M_m(1 S)}{M_b(1 D)-M_b(1 S)}}$   &  1.27      &  1.25     &  1.33   \\

\toprule
\end{tabular}
\end{table}
\end{center}  
\subsection{Symmetry Multiplets}
The quark model states for the $\Xi_{cc}$, $\Xi_{bc}$, and the $\Xi_{bb}$ from the spinless RP model are arranged into symmetry multiplets in table \ref{multipletsSpinless}, and the results from the full model \cite{RobertsPervin} are in table \ref{multipletsSpin}.  The results for the full model and the spinless model for flavor-singlet $\Xi^\prime_{bc}$ states are in table \ref{multipletsOdd}. Tables \ref{multipletsSpin} and \ref{multipletsOdd} show that the inclusion of spin dependent interactions lift the degeneracy of the states by less than 75 MeV, and these splittings decrease as the heavy diquark's mass increases.  One interesting feature of the model is that there are degeneracies in the spectrum in addition to those predicted by the decoupling of the heavy diquark's spin.  These additional degeneracies arise due to the fact that the RP model does not include a fully microscopic spin-orbit interaction.  For instance, in table \ref{multipletsSpin} the $n_d^{2s_d+1}(l_\rho)_{J_d}=1^{3}D_1$, $1^{3}D_2$,  and $1^{3}D_3$ multiplets are nearly degenerate.  An appropriate spin-orbit interaction would split these states, but this would be suppressed by the masses of the heavy quarks.  There are also additional degeneracies among the $n_d^{2s_d+1}(l_\rho)_{J_d}$, $n_\lambda(l_\lambda)_{J_l}=1^{3}S_1$, $1P_\frac{1}{2}$ and $1^{3}S_1$, $1P_\frac{3}{2}$  multiplets as well as the $1^{1}P_1$, $1P_\frac{1}{2}$ and $1^{1}P_1$, $1P_\frac{3}{2}$ multiplets.  These degeneracies would be lifted if the model included a spin-orbit interaction which couples the orbital angular momentum of the heavy-light system to the spin of the light quark, such as an operator proportional to $\vec{l}_{\lambda} \cdot \vec{s}_l$.  Such an interaction is not expected to be suppressed in the heavy quark limit.
\begin{center}
\begin{table}[h]
\caption{Symmetry multiplets for the RP model for flavor-triplet states with spin interactions removed.  The decoupling of the diquark total angular momentum predicts that the states within a multiplet will be degenerate.
\label{multipletsSpinless}}
\vspace{2mm}
\renewcommand{\arraystretch}{1.0}
\begin{tabular}{c c c c c }
\hline \toprule
Multiplet                         & $J^P$                       & \multicolumn{3}{c }{Mass (GeV)} \\ 
\hspace{1pt} $n_d^{2s_d+1}(l_\rho)_{J_d}$, $n_\lambda(l_\lambda)_{J_l}$ \hspace{1pt} 
                                        & $(|J_d-J_l|, ... ,J_d+J_l)$ & $\Xi_{cc}$ & $\Xi_{bc}$ & $\Xi_{bb}$\\ \hline
$1^{3}S_1$, $1S_\frac{1}{2}$             & $(\hlf^+,\thlf^+)$          & (3.724, 3.724) & (7.045, 7.045) & (10.341, 10.341) \\ 
$1^{1}P_1$, $1S_\frac{1}{2}$             & $(\hlf^-,\thlf^-)$          & (3.912, 3.912) & (7.212, 7.212) & (10.470, 10.470) \\ 
$2^{3}S_1$, $1S_\frac{1}{2}$             & $(\hlf^+,\thlf^+)$          & (4.056, 4.056) & (7.339, 7.339) & (10.564, 10.564) \\ 
$1^{3}D_1$, $1S_\frac{1}{2}$             & $(\hlf^+,\thlf^+)$          & (4.078, 4.078\footnotemark[1]) & (7.358, 7.358\footnotemark[1]) & (10.579, 10.579\footnotemark[1]) \\
$1^{3}D_2$, $1S_\frac{1}{2}$             & $(\thlf^+,\fhlf^+)$         & (4.078\footnotemark[1], 4.079) & (7.358\footnotemark[1], 7.360) & (10.579\footnotemark[1], 10.581) \\
$1^{3}D_3$, $1S_\frac{1}{2}$             & $(\fhlf^+,\shlf^+)$         & (4.079, 4.078) & (7.360, 7.358) & (10.582, 10.579) \\
$1^{3}S_1$, $1P_\frac{1}{2}$             & $(\hlf^-,\thlf^-)$          & (4.073, 4.073\footnotemark[2]) & (7.391, 7.391\footnotemark[2]) & (10.691, 10.691\footnotemark[2]) \\ 
$1^{3}S_1$, $1P_\frac{3}{2}$            & $(\hlf^-,\thlf^-,\fhlf^-)$   & (4.073, 4.073\footnotemark[2], 4.073) & (7.391, 7.391\footnotemark[2], 7.391) & (10.691, 10.691\footnotemark[2], 10.691) \\ 
$1^{1}P_1$, $1P_\frac{1}{2}$             & $(\hlf^+,\thlf^+)$          & (4.259\footnotemark[3], 4.249) & (7.557\footnotemark[3], 7.548) & (10.818\footnotemark[3], 10.813) \\ 
$1^{1}P_1$, $1P_\frac{3}{2}$            & $(\hlf^+,\thlf^+,\fhlf^+)$   & (4.235\footnotemark[3], 4.259, 4.247) & (7.536\footnotemark[3], 7.557, 7.546) & (10.804\footnotemark[3], 10.818, 10.809) \\
$1^{3}S_1$, $2S_\frac{1}{2}$          & $(\hlf^+,\thlf^+)$             & (4.346, 4.346) & (7.662, 7.662) & (10.964, 10.963) \\ 
$1^{3}S_1$, $1D_\frac{3}{2}$          & $(\hlf^+,\thlf^+,\fhlf^+)$     & (4.392, 4.392, 4.385\footnotemark[4]) & (7.709, 7.709, 7.700\footnotemark[4]) & (11.013,  11.013, 11.001\footnotemark[4]) \\ 
$1^{3}S_1$, $1D_\frac{5}{2}$          & $(\thlf^+,\fhlf^+, \shlf^+)$   & (4.392, 4.385\footnotemark[4], 4.392) & (7.709, 7.700\footnotemark[4], 7.709) & (11.013,  11.001\footnotemark[4], 11.013) \vspace{.5mm} \\  
\toprule
\footnotetext[1]{$^{\text{bcd}}$These states mix significantly.}
\end{tabular}
\end{table}
\end{center}

The energy splittings due to pairwise mixing (section \ref{pairmix}) are small in the RP model.  For instance, the mass splittings due to the mixing of states with $l_\rho=0$, $l_\lambda=2$ and $l_\rho=2$, $l_\lambda=0$ are not noticeable in the spectra.  The mixing of states due to this is also quite small and vanishes as the heavy quark mass increases (although there is no {\it a priori} reason to expect it to).  States that are orbitally excited in both the heavy diquark and the light degrees of freedom are also subject to pairwise mixing in the $J_d$ and $J_l$ quantum numbers.  This is the reason for the broken degeneracies of the $n_d^{2s_d+1}(l_\rho)$, $n_\lambda(l_\lambda)=1^{1}P$, $1P$ multiplets in table \ref{multipletsSpinless} and the $1^{3}P$, $1P$ multiplets in the ``Spinless'' column of table \ref{multipletsOdd}.  All of these splittings are small (less than 25 MeV), but the mixing associated with them is substantial, particularly for the flavor-singlet states.

\begin{center}
\begin{table}[!h]
\caption{Symmetry multiplets for the RP model for flavor-triplet states including spin dependent interactions.  The decoupling of the diquark total angular momentum predicts that the states within a multiplet will be approximately degenerate.
\label{multipletsSpin}}
\vspace{2mm}
\renewcommand{\arraystretch}{1.0}
\begin{tabular}{c c c c c }
\hline \toprule
Multiplet                          & $J^P$                        & \multicolumn{3}{c }{Mass (GeV)} \\ 
\hspace{1pt} $n_d^{2s_d+1}(l_\rho)_{J_d}$, $n_\lambda(l_\lambda)_{J_l}$ \hspace{1pt} & $(|J_d-J_l|, ... ,J_d+J_l) $ & $\Xi_{cc}$ & $\Xi_{bc}$ & $\Xi_{bb}$\\ \hline
$1^{3}S_1$, $1S_\frac{1}{2}$              & $(\hlf^+,\thlf^+)$           & (3.678, 3.752) & (7.014, 7.064) & (10.322, 10.352) \\ 
$1^{1}P_1$, $1S_\frac{1}{2}$              & $(\hlf^-,\thlf^-)$           & (3.911, 3.917) & (7.212, 7.214) & (10.470, 10.470) \\ 
$2^{3}S_1$, $1S_\frac{1}{2}$              & $(\hlf^+,\thlf^+)$           & (4.030, 4.078) & (7.321, 7.353) & (10.551, 10.574) \\ 
$1^{3}D_1$, $1S_\frac{1}{2}$              & $(\hlf^+,\thlf^+)$           & (4.098, 4.045\footnotemark[1]) & (7.372, 7.337\footnotemark[1]) & (10.589, 10.564\footnotemark[1]) \\
$1^{3}D_2$, $1S_\frac{1}{2}$              & $(\thlf^+,\fhlf^+)$          & (4.094\footnotemark[1], 4.092) & (7.369\footnotemark[1], 7.368) & (10.587\footnotemark[1], 10.588) \\
$1^{3}D_3$, $1S_\frac{1}{2}$              & $(\fhlf^+,\shlf^+)$          & (4.048, 4.095) & (7.340, 7.370) & (10.568, 10.588) \\
$1^{3}S_1$, $1P_\frac{1}{2}$              & $(\hlf^-,\thlf^-)$           & (4.081, 4.077) & (7.397, 7.392) & (10.694, 10.691) \\ 
$1^{3}S_1$, $1P_\frac{3}{2}$             & $(\hlf^-,\thlf^-,\fhlf^-)$   & (4.073, 4.079, 4.089) & (7.390, 7.394, 7.399) & (10.691, 10.692, 10.695) \\ 
$1^{1}P_1$, $1P_\frac{1}{2}$              & $(\hlf^+,\thlf^+)$           & (4.257\footnotemark[2], 4.253) & (7.555\footnotemark[2], 7.549) & (10.817\footnotemark[2], 10.812) \\ 
$1^{1}P_1$, $1P_\frac{3}{2}$             & $(\hlf^+,\thlf^+,\fhlf^+)$   & (4.230\footnotemark[2], 4.261, 4.259) & (7.534\footnotemark[2], 7.557, 7.549) & (10.802\footnotemark[2], 10.818, 10.811) \\
$1^{3}S_1$, $2S_\frac{1}{2}$          & $(\hlf^+,\thlf^+)$             & (4.311, 4.368) & (7.634, 7.676) & (10.940, 10.972) \\ 
$1^{3}S_1$, $1D_\frac{3}{2}$          & $(\hlf^+,\thlf^+,\fhlf^+)$     & (4.394, 4.394, 4.387) & (7.709, 7.708, 7.701) & (11.011,  11.011, 11.002) \\ 
$1^{3}S_1$, $1D_\frac{5}{2}$          & $(\thlf^+,\fhlf^+,\shlf^+)$    & (4.391, 4.388, 4.393) & (7.706, 7.702, 7.708) & (11.011,  11.002, 11.011) \vspace{.5mm} \\ 
\toprule
\footnotetext[1]{$^{\text{b}}$These states mix significantly.}
\end{tabular}
\end{table}
\end{center}
\begin{center}
\begin{table}[!h]
\caption{Symmetry multiplets for the RP model for flavor-singlet states.  The decoupling of the diquark total angular momentum predicts that the states within a multiplet will be approximately degenerate.
\label{multipletsOdd}}
\vspace{2mm}
\renewcommand{\arraystretch}{1.0}
\begin{tabular}{c c c c }
\hline \toprule
Multiplet                   & $J^P$                        & \multicolumn{2}{c }{$\Xi^\prime_{bc}$ Mass (GeV)} \\ 
\hspace{1pt} $n_d^{2s_d+1}(l_\rho)_{J_d}$, $n_\lambda(l_\lambda)_{J_l}$ \hspace{1pt} & $(|J_d-J_l|, ... ,J_d+J_l) $ & Spinless & Spin \\ \hline
$1^{1}S_0$, $1S_\frac{1}{2}$       & $(\hlf^+)$                   & (7.053) & (7.037)  \\ 
$1^{3}P_0$, $1S_\frac{1}{2}$       & $(\hlf^-)$                   & (7.219\footnotemark[1]) & (7.230\footnotemark[1])  \\ 
$1^{3}P_1$, $1S_\frac{1}{2}$       & $(\hlf^-,\thlf^-)$           & (7.219\footnotemark[1], 7.219) & (7.199\footnotemark[1], 7.228)  \\ 
$1^{3}P_2$, $1S_\frac{1}{2}$       & $(\thlf^-,\fhlf^-)$          & (7.219, 7.217) & (7.201, 7.265)  \\ 
$2^{1}S_0$, $1S_\frac{1}{2}$       & $(\hlf^+)$                   & (7.349) & (7.333)  \\ 
$1^{1}D_2$, $1S_\frac{1}{2}$       & $(\thlf^+,\fhlf^+)$          & (7.367, 7.367) & (7.361, 7.367)  \\
$1^{1}S_0$, $1P_\frac{1}{2}$       & $(\hlf^-)$                   & (7.398) & (7.388) \\ 
$1^{1}S_0$, $1P_\frac{3}{2}$      & $(\thlf^-)$                  & (7.398) & (7.390) \\ 
$1^{3}P_0$, $1P_\frac{1}{2}$       & $(\hlf^+)$                   & (7.553\footnotemark[2]) & (7.548\footnotemark[2]) \\ 
$1^{3}P_0$, $1P_\frac{3}{2}$      & $(\thlf^+)$                  & (7.561) & (7.546\footnotemark[3]) \\ 
$1^{3}P_1$, $1P_\frac{1}{2}$       & $(\hlf^+,\thlf^+)$           & (7.562\footnotemark[2], 7.553\footnotemark[3]) 
                                                           & (7.552\footnotemark[2], 7.556\footnotemark[3])  \\ 
$1^{3}P_1$, $1P_\frac{3}{2}$      & $(\hlf^+,\thlf^+,\fhlf^+)$   & (7.562\footnotemark[2], 7.553\footnotemark[3], 7.553\footnotemark[4]) 
                                                           & (7.555\footnotemark[2], 7.548\footnotemark[3], 7.553\footnotemark[4])  \\ 
$1^{3}P_2$, $1P_\frac{1}{2}$       & $(\thlf^+,\fhlf^+)$          & (7.541\footnotemark[3], 7.553\footnotemark[4]) 
                                                           & (7.563\footnotemark[3], 7.548\footnotemark[4])  \\ 
$1^{3}P_2$, $1P_\frac{3}{2}$      & $(\hlf^+,\thlf^+,\fhlf^+,\shlf^+)$   & (7.541\footnotemark[2], 7.561\footnotemark[3], 7.561\footnotemark[4], 7.553) 
                                                                    & (7.519\footnotemark[2], 7.562\footnotemark[3], 7.559\footnotemark[4], 7.557)  \\ 
$1^{1}S_0$, $2S_\frac{1}{2}$       & $(\hlf^+)$                   & (7.661) & (7.645)  \\ 
$1^{1}S_0$, $1D_\frac{3}{2}$       & $(\thlf^+)$                  & (7.707) & (7.709)  \\ 
$1^{1}S_0$, $1D_\frac{5}{2}$       & $(\fhlf^+)$                  & (7.707) & (7.689)  \vspace{.5mm} \\ 
\toprule
\footnotetext[1]{$^{\text{bcd}}$These states mix significantly.}
\end{tabular}
\end{table}
\end{center}
\subsection{Diquark Excitation Symmetry}
The extended superflavor symmetry (Section \ref{Superflav}), dictates that the DHB spectrum will exhibit a repeating pattern corresponding to the excitations of the light degrees of freedom; each occurrence of this pattern will be ``shifted" by the mass of the heavy diquark.  Table \ref{SuperFlavTablea} shows the energy required to excite the light degrees of freedom in flavor-triplet DHBs in the spinless RP model for multiplets with two different diquark excitations: the ground state and the $n_d^{2s_d+1}(l_\rho)_{J_d}$=$1^1P_1$ multiplets.  Table \ref{FlavTableOdd} shows analogous splittings for flavor-singlet states, for the ground state and the $^3P_{J_d}$ diquarks.  The excitation symmetry requires that each entry in the first row of table \ref{SuperFlavTablea} should be equal to the entry beneath it on the second row, and each entry on the third row should be identical to the one beneath it on the fourth row.  Likewise, the entries in the first four rows of table \ref{FlavTableOdd} should be equal, as should the entries in rows five to eight.  This symmetry should be broken by this particular model because of pair-wise interactions, but all of the excitation energies that should be identical agree to within 14 MeV for the flavor-triplet states, and 16 MeV for the flavor-singlet states.  We can therefore conclude that the effects of pair-wise forces on this aspect of the spectrum are small.
\begin{center}
\begin{table}[h]
\caption{Energy required to excite the light degrees of freedom of a flavor-triplet DHB from the ground state, $n_\lambda(l_\lambda)_{{J_l}^{\pi_l}}=1S_{\hlf^+}$, to the excited state listed in the second column, for different states of the heavy diquark, listed in the first column.  We have averaged over the masses in each multiplet.  This excitation energy should be independent of the spin, parity, and flavor of the heavy diquark, so that the energies in the first row should be the same as the corresponding ones in the second row, and the entries in the third row should equal those in the fourth row. These excitation energies should also be independent of the flavor of the quarks that comprise the heavy diquark, so that all entries on a particular row should be equal. This is seen not only for the first four rows, but also for the last three rows.
\label{SuperFlavTablea}}
\vspace{2mm}
\renewcommand{\arraystretch}{1.0}
\begin{tabular}{ccccccc}
\hline \toprule
Diquark state & & Upper Multiplet  & $\hphantom{aaa}$  Lower Multiplet  $\hphantom{a}$ &  \multicolumn{3}{c}{$m^*-m_0$ (MeV)}\\\cline{5-7}
$n_d^{2s_d+1}(l_\rho)_{J_d}$& $\hphantom{ae}n_\lambda(l_\lambda)_{J_l}\hphantom{ab}$& ($m^*$)   & $\hphantom{\hlf^+}$ ($m_0$)  & $\hphantom{a}\Xi_{cc}\hphantom{a}$ & $\hphantom{a}\Xi_{bc}\hphantom{a}$ & $\hphantom{a}\Xi_{bb}\hphantom{a}$ \\ \hline 

$1^3S_1$& $1P_\hlf$  &  $\hphantom{\hlf^+,}(\hlf^-,\thlf^-)$ & $\hphantom{\hlf^+,}(\hlf^+,\thlf^+)$ & 349 & 346 & 350 \\ 
$1^1P_1$& $1P_\hlf$  &   $\hphantom{\hlf^+,}(\hlf^+,\thlf^+)$    & $\hphantom{\hlf^+,}(\hlf^-,\thlf^-)$ & 342 & 341 & 346 \\ 
$1^3S_1$& $1P_\thlf$ &   $(\hlf^-,\thlf^-,\fhlf^-)$    &$\hphantom{\hlf^+,}(\hlf^+,\thlf^+)$  & 349 & 346 & 350 \\ 
$1^1P_1$& $1P_\thlf$ & $(\hlf^+,\thlf^+,\fhlf^+)$      & $\hphantom{\hlf^+,}(\hlf^-,\thlf^-)$ & 335 & 334 & 340  \\ \\
$1^3S_1$& $1D_\thlf$ & $(\hlf^+,\thlf^+,\fhlf^+)$      & $\hphantom{\hlf^+,}(\hlf^+,\thlf^+)$ & 666 & 661 & 668 \\
$1^3S_1$& $1D_\fhlf$ & $(\thlf^+,\fhlf^+,\shlf^+)$      & $\hphantom{\hlf^+,}(\hlf^+,\thlf^+)$ & 665 & 660 & 666 \\
$1^3S_1$& $2S_\hlf$  & $\hphantom{\hlf^+,}(\hlf^+,\thlf^+)$      & $\hphantom{\hlf^+,}(\hlf^+,\thlf^+)$ & 622 & 617 & 623\\ 

\toprule
\end{tabular}
\end{table}
\end{center}
\begin{center}
\begin{table}[!h]
\caption{Energy required to excite the light degrees of freedom of the flavor singlet DHB from the ground state, $n_\lambda(l_\lambda)_{{J_l}^{\pi_l}}=1S_{\hlf^+}$, to the excited state listed in the second column, for different states of the heavy diquark, listed in the first column.  We have averaged over the masses in each multiplet.  The excitation energies on the first four rows should be equal, as should those on rows five to eight. 
\label{FlavTableOdd}}
\vspace{2mm}
\renewcommand{\arraystretch}{1.0}
\begin{tabular}{ccccc}
\hline \toprule
Diquark state & & $\hphantom{\hlf^+}$ Upper Multiplet  & $\hphantom{\hlf^+}$ Lower Multiplet & $m^*-m_0$ (MeV) \\
\hspace{1pt}$n_d^{2s_d+1}(l_\rho)_{J_d}$& $\hphantom{ae}n_\lambda(l_\lambda)_{J_l}\hphantom{ab}$& $\hphantom{\hlf^+}$($m^*$)    & $\hphantom{\hlf^+}$($m_0$)  & $\hphantom{aaaaaaa}\Xi_{bc}\hphantom{aaaaaaa}$ \\ \hline 

$1^1S_0$& $1P_\hlf$  & $\hphantom{\hlf^+,}(\hlf^-)$ &$\hphantom{\hlf^+,}(\hlf^+)$  & 345  \\ 
$1^3P_0$& $1P_\hlf$  &  $\hphantom{\hlf^+,}(\hlf^+)$& $\hphantom{\hlf^+,}(\hlf^-)$ & 334  \\ 
$1^3P_1$& $1P_\hlf$  & $\hphantom{\hlf^+,}(\hlf^+,\thlf^+)$& $\hphantom{\hlf^+,}(\hlf^-,\thlf^-)$ & 339  \\ 
$1^3P_2$& $1P_\hlf$  &  $\hphantom{\hlf^+,}(\thlf^+,\fhlf^+)$ &$\hphantom{\hlf^+,}(\thlf^-,\fhlf^-)$ & 329  \\ 
$1^1S_0$& $1P_\thlf$ &  $\hphantom{\hlf^+,}(\thlf^-)$ &$\hphantom{\hlf^+,}(\hlf^+)$& 345  \\ 
$1^3P_0$& $1P_\thlf$ &  $\hphantom{\hlf^+,}(\thlf^+)$ &$\hphantom{\hlf^+,}(\hlf^-)$& 342    \\ 
$1^3P_1$& $1P_\thlf$ &  $\hphantom{\hlf^+,}(\hlf^+,\thlf^+,\fhlf^+)$& $\hphantom{\hlf^+,}(\hlf^-,\thlf^-)$ & 337  \\ 
$1^3P_2$& $1P_\thlf$ &  $\hphantom{\hlf^+,}(\hlf^+,\thlf^+,\fhlf^+,\shlf^+)$ &$\hphantom{\hlf^+,}(\thlf^-,\fhlf^-)$ & 336   \\ \\
$1^1S_0$& $1D_\thlf$ & $\hphantom{\hlf^+,}(\thlf^+)$ & $\hphantom{\hlf^+,}(\hlf^+)$ & 654  \\
$1^1S_0$& $1D_\fhlf$ & $\hphantom{\hlf^+,}(\fhlf^+)$& $\hphantom{\hlf^+,}(\hlf^+)$  & 654  \\
$1^1S_0$& $2S_\hlf$  & $\hphantom{\hlf^+,}(\hlf^+)$& $\hphantom{\hlf^+,}(\hlf^+)$ & 608  \\ 
\toprule
\end{tabular}
\end{table}
\end{center}
\subsection{Diquark Flavor Symmetry}
The heavy diquark flavor symmetry dictates that the energy required to excite the light degrees of freedom should not depend on the mass of the heavy diquark.  This can be seen for flavor-triplet states in table \ref{SuperFlavTablea}, where the entries on each row are approximately equal: in the spinless Roberts-Pervin model these relations work out within 7 MeV.  We wish to re-emphasize that this result has emerged from a model which does not explicitly factorize in the heavy quark limit and does not assume anything about the size of the heavy diquark.  We can also see if the energy splittings depend on the statistics of the heavy diquark.  Table \ref{FlavTableOdd} shows the same energy splittings as table \ref{SuperFlavTablea} for the $\Xi^\prime_{bc}$ states. As an example, the energy required to excite the light degrees of freedom from the $n_\lambda(l_\lambda)_{J_l}=1S_\hlf$ multiplet to the $1P_\hlf$ multiplets in table \ref{SuperFlavTablea} (rows 1 and 2) should match the excitation energies in table \ref{FlavTableOdd} (rows 1 through 4).  Comparing the splittings for the $\Xi^\prime_{bc}$ states with the splittings for the $\Xi_{bc}$ states we see that the expected relation holds within 21 MeV.

\subsection{Superflavor Symmetry} \label{SupFlavResults}

In this section we use the superflavor symmetry to compare the spectra of singly heavy mesons to DHBs \cite{SavageSpectrum}.  We do not extensively compare these two types of hadrons here, but rather, we simply compare the energy required to orbitally excite the light degrees of freedom of a DHB to the analogous splitting in single heavy mesons.  There are not enough experimental data to compare with $J_l^{\pi_l}=\hlf^-$ states, but there are experimental data for the $J_l^{\pi_l}=\thlf^-$ states.  The flavor symmetry of HQS predicts that these splittings should be the same for $D$ mesons and $B$ mesons (neglecting $1/M_Q$ corrections), and the superflavor symmetry dictates that this excitation energy should be identical to the third and fourth rows of splittings for DHBs in table \ref{SuperFlavTablea}.  Table \ref{FlavTableThlfm} contains these splittings for heavy mesons.  This treatment seems to show that the model of Ref. \cite{RobertsPervin} underestimates the size of the orbital splitting of the light degrees of freedom.  We also see an indication that the charm quark's mass may still be too light for the flavor symmetry to emerge to a high degree of accuracy.  Nevertheless, we can use these results to estimate this splitting to be on the order of 400 MeV for DHBs.

In the RP model there is a separation in mass between states that involve excitations of the heavy diquark and states that involve excitations of the light degrees of freedom.  For instance, the difference in energy between the ground state and the first orbital excitation of the heavy diquark is about 130 MeV for the $\Xi_{bb}$.  In contrast, the difference in energy between the ground state and the first orbital excitation of the light quark is about 350 MeV.  Furthermore, the mixing between states with different orbital excitations of the diquark is small.  We can determine if this separation holds using the estimate of 230 MeV for the orbital splitting of heavy diquarks from section \ref{secMBrat} and the estimate of 400 MeV for the energy required to excite the light degrees of freedom that we presented above.  This establishes a hierarchy of energy scales:  $\Lambda_d \ll \mu_{Q_1 Q_2}$, $\Lambda_l \ll M_d \sim M_{Q_1}+M_{Q_2}$, and $\Lambda_d < \Lambda_l$;  $\Lambda_d$ is the energy required to excite the heavy diquark, $\Lambda_l \sim \Lambda_{QCD}$ is the energy required to excite the light degrees of freedom, $M_d$ is the mass of the heavy diquark, and $\mu_{{Q_1} {Q_2}}$ is the reduced mass of the heavy diquark subsystem.
\begin{center} 
\begin{table}[h]
\caption{Spin averaged masses and the energy (MeV) required to excite the light degrees of freedom of singly heavy mesons from the ground state $J_l^{\pi_l}=\hlf^+$ state to the $J_l^{\pi_l}=\thlf^-$ state \cite{PDG}.  We have also included results for Kaons.
\label{FlavTableThlfm}}
\vspace{2mm}
\begin{tabular}{cccc}
\hline \toprule
  Flavor  &   $ m_{(0^-,1^-)} $&$ m_{(1^+,2^+)} $&$\Delta m$ \\ \hline
   $K$&       793.15 $\pm$ 0.20 & 1368.0 $\pm$ 2.5   & 574.9 $\pm$ 2.5 \\

   $D$&        1971.43 $\pm$ 0.13 &\hspace{6pt} 2446.46 $\pm$ 0.52 \hspace{6pt}& 475.04 $\pm$ 0.53 \\

   $B$&      5313.70 $\pm$ 0.40 & 5735.7 $\pm$ 2.9   & 422.0 $\pm$ 2.9    \\
\toprule
\end{tabular}

\end{table}
\end{center}

\subsection{Strange Diquarks}

In many quark models, the strange quark's mass is similar in magnitude to $\Lambda_{\rm QCD}$. This should be much too light for this quark to be considered heavy, but features in the calculated spectrum of ref. \cite{RobertsPervin} emerge that are very easily understood if this odd approximation is made. If the strange quark is treated as heavy, the flavor multiplets of section \ref{flavormultiplets} need to be modified. Instead of the broken SU(2) of that section, we now consider a (badly) broken ${\text SU(3)}_{\text heavy}$ flavor symmetry.  There will now be a sextet of symmetric states
\begin{equation}
 \Xi=ssq,\,\,\,\,\,\ \Xi_{cc}=ccq,\,\,\,\,\,\Xi_{bb}=bbq,\,\,\,\,\Xi_{c}=\frac{1}{\sqrt{2}}\left(sc+cs\right)q,\,\,\,\,\Xi_{b}=\frac{1}{\sqrt{2}}\left(sb+bs\right)q,\,\,\,\,\Xi_{bc}=\frac{1}{\sqrt{2}}\left(cb+bc\right)q,
\end{equation}
and an antitriplet of antisymmetric states
\begin{equation}
\Xi^\prime_{c}=\frac{1}{\sqrt{2}}\left(sc-cs\right)q,\,\,\,\,\Xi^\prime_{b}=\frac{1}{\sqrt{2}}\left(sb-bs\right)q,\,\,\,\,\Xi^\prime_{bc}=\frac{1}{\sqrt{2}}\left(cb-bc\right)q.
\end{equation}
For these states, there are only two choices for the light quark, $q$, namely $u$ and $d$.
These multiplets should not be confused with the multiplets in which singly-heavy baryons are placed, with the strange quark taken as one of the {\em light} triplet of quarks. The ${\text SU(3)}_{\text light}$ sextet then contains the states
\begin{equation}
 \Sigma_Q=uuQ,\,\, ddQ,\,\,\frac{1}{\sqrt{2}}\left(ud+du\right)Q;\,\,\,\Xi^\prime_{Q}=\frac{1}{\sqrt{2}}\left(su+us\right)Q, \,\,\frac{1}{\sqrt{2}}\left(sd+ds\right)Q;\,\,\,\,\Omega_{Q}=ssQ,
\end{equation}
and the corresponding antitriplet consists of
\begin{equation}
\Lambda_Q=\frac{1}{\sqrt{2}}\left(ud-du\right)Q,\,\,\,\,\Xi_{Q}=\frac{1}{\sqrt{2}}\left(us-su\right)Q,\,\,\frac{1}{\sqrt{2}}\left(ds-sd)\right)Q. 
\end{equation}
In these multiplets, the heavy quark $Q$ may be either $b$ or $c$.

In the following subsections we examine the consequences of treating the strange quark as a heavy quark in doubly heavy systems. The symmetries and systematics we have discussed in the preceding sections should not be expected to work well when applied to baryons in which the strange quark is treated as heavy. Nevertheless, examining such a limit may provide clues to the understanding of the spectra of such baryons, particularly the $\Xi$s. It may also provide some information on how badly the symmetries get broken as the masses of the heavy quarks decrease. 

\subsubsection{Meson-Meson Ratios}

We begin this investigation with meson-meson ratios, eqn. (\ref{DHMratio}).  These ratios allow us to extract the effective energy scaling index, $d$, for the region of reduced masses involving strange diquarks. For strangeonia there is an appreciable amount of flavor mixing, and calculations of this mixing are inherently model dependent.  Therefore, we rely on phenomenological models to estimate the mixing angles and to extract the masses of pure $s \overline{s}$ states.  We treat three states as unmixed $s \overline{s}$ mesons:  $\phi (1020)$, $h_1(1380)$, and the $f^\prime_2(1525)$.  For the $\eta$-$\eta^\prime$ mixing we use the mixing angle of $42^\circ$ due to Thomas \cite {Thomas}; for the $1^3P_0$ $s \overline{s}$ state we use the mass of $1682 \pm 4$ MeV determined by Close and Zhao \cite{CloseZhao}; and for the $f_1(1285)$-$f_1(1420)$ mixing angle we use the value of $21 ^\circ \pm 5^\circ$ due to a method by Close and Kirk \cite{CloseKirk} updated with more recent experimental data \cite{Dudek}.  These mixing angles are similar to those found in a lattice calculation \cite{Dudek}.  At this time, there are only enough experimental data to calculate the spin-averaged masses of the ground state and the first orbitally excited state of $D_s$ mesons.  We calculate the orbital splittings of these mesons in table \ref{sMeson-MesonTable}.  The errors reported in these results only reflect the experimental uncertainties, and the theoretical error due to the choice of phenomenological descriptions is left undetermined.
\begin{center} 
\begin{table}[h]
\caption{Spin averaged mass splittings for $D_s$ mesons and strangeonium from experiment \cite{PDG}.  The energy scaling rule, eqn. (\ref{Escale}), predicts that the ratio of these splittings should be $\left(\frac{\mu_{ss}}{\mu_{sc}}\right)^{d-1}$.  We also extract the energy scaling parameter, $d$, from the ratio using the heavy quark masses from the RP model \cite{RobertsPervin}.
\label{sMeson-MesonTable}}
\vspace{2mm}
\begin{tabular}{ccccc}
\hline \toprule
\hphantom{ab} Splitting \hphantom{ab}&\hphantom{ab}$\Delta m_{s \overline{s}}$ (MeV)\hphantom{ab}&\hphantom{ab}$\Delta m_{c \overline{s}}$ (MeV)\hphantom{ab}&\hphantom{ab}$\Delta m_{s \overline{s}}/\Delta m_{c \overline{s}}$ \hphantom{ab}&   $d$ \\ \hline
$M(1P)-M(1S)$  &  515.11 $\pm$ 7.23 &    437.42 $\pm$ 0.56           &          1.178 $\pm$ 0.017      &  0.617 $\pm$ 0.076 \\
\toprule
\end{tabular}
\end{table}
\end{center}

The result for the scaling index $d$ is close to the value that one would expect for a purely linear confining potential ($d=\frac{2}{3}$).  This suggests that the nonperturbative nature of QCD has become more important for this region of reduced masses.  We note that, unlike the ratios obtained in section \ref{mesonmeson}, the value of $d$ extracted depends on the quark model masses used, and the value of 0.617 is obtained using the quark masses of the RP model. 
 This is significantly different from the scaling index for bottomonia and charmonia, where $d \approx 1$ and the quark mass dependence vanishes.  It is interesting to note that the orbital splitting of the $c \overline{s}$ system is nearly the same as the analogous splitting in $c \overline{c}$ and $b \overline{b}$ systems (see table \ref{Meson-MesonTable}).  This suggests that $d \approx 1$ scaling holds reasonably well for a range of reduced masses from that of the $b \overline{b}$ system down to that of the $c \overline{s}$ system, and then approaches $d=\frac{2}{3}$ scaling appropriate to a linear potential for $s \overline{s}$ systems.

\subsubsection{Meson-Baryon Ratios}

With the value of $d=0.617 \pm 0.076$, we expect the ratio of the orbital splitting in the $s \overline{s}$ meson to that of the heavy diquark in a $\Xi$ baryon to be $1.534 \pm 0.035$.  For the $\Xi_c$ baryons, there are not enough experimental data to fully determine the orbital excitation energy of the $sc$ diquark.  The PDG \cite{PDG} results show all three ground states, along with two low-lying negative parity states which might be interpreted as containing orbitally excited heavy diquarks.  There should be five additional negative parity states associated with orbital heavy diquark excitations, but those states are not yet identifiable.  For the $\Xi$, there are only 2 low-lying, negative-parity states associated with an orbitally excited heavy diquark, and these two states are readily identifiable.  Based on the RP model predictions, we expect the two lowest excited states to be orbital excitations of the heavy diquark; the $\Xi(1820)$ has quantum numbers $J^P=\thlf^-$, and we expect the $\Xi(1690)$ to have quantum numbers $J^P=\hlf^-$ based on general features of quark models. Treating these states as pure diquark excitations, we calculate the orbital splittings of the $\Xi$ baryon and the expected splitting derived from the meson-baryon ratios (eqn. \ref{MBratio}) and the value of the scaling index extracted from strangeonium and $D_s$ meson orbital splittings (table \ref{sMeson-MesonTable}).  The predictions from this application of the supercolor symmetry are in table \ref{sMeson-BaryonTable} along with the spin-averaged prediction from the RP model.
 
\begin{center} 
\begin{table}[h]
\caption{$\Xi$ and $\Xi_c$ heavy diquark orbital splittings.  The Supercolor Symmetry predictions come from meson-baryon ratios (eqn. \ref{MBratio}).  We have used the estimate for $d$ obtained from $s \overline{s}$ to $c \overline{s}$ meson-meson ratios (table \ref{sMeson-MesonTable}) in rows 1 and 3 and the estimate for $d$ from $c \overline{s}$ to $c \overline{c}$ meson-meson ratios in the last row.  These estimates depend on the quark masses of the RP model.  The second row uses a value of $d=2/3$, which one would expect from a nonrelativistic treatment with a linear confining potential.
\label{sMeson-BaryonTable}}
\vspace{2mm}
\renewcommand{\arraystretch}{1}
\begin{tabular}{ccccccc}
\hline \toprule
 Baryon & Splitting  & $d$ &\hspace{3pt}Supercolor Symmetry (MeV)\hspace{3pt}&\hspace{3pt}RP Model\hspace{3pt} &\hspace{3pt}Experiment (MeV)\hspace{3pt}\\ \hline
\multirow{2}{*}{$\Xi \hphantom{_c}$}  &\multirow{2}{*}{${\displaystyle M(1 P)-M(1 S)}$}      & 0.617 $\pm$ 0.076 &  335.9 $\pm$ 9.0          & \multirow{2}{*}{292.7} &  \multirow{2}{*}{317.0 $\pm$ 4.7}  \\
       & & 0.667             &  324.5 $\pm$ 4.6          &  &    \\\\
\multirow{2}{*}{$\Xi_c$}&\multirow{2}{*}{ ${\displaystyle M(1 P)-M(1 S)}$ }  & 0.617 $\pm$ 0.076 &   285.2 $\pm$ 6.5          & \multirow{2}{*}{254.9} &  \multirow{2}{*}{--}  \\
       & & 1.059 $\pm$ 0.004 &  209.9 $\pm$ 0.4          &  &    \\
\toprule
\end{tabular}
\end{table}
\end{center}  

The experimental splitting and the splitting calculated using the supercolor symmetry are within 6\% of each other.  This agreement arises despite a number of possible issues with this treatment:  (a) this prediction assumes non-relativistic kinematics for the quarks in the $ss$ system;  (b)  the expected value for the splitting depends on the $s$ and $c$ quark masses used; (c) the solution to the energy scaling equation (eqn. \ref{Escale}) from which we derived this result is only an approximation from which the value of $d$ is chosen so that the solution reproduces experimental data for a finite range of reduced diquark masses;  (d)  we have ignored the mixing of heavy diquark excitations with excitations of the light degrees of freedom due to tensor interactions;  and (e) the results for the $\Xi$ splitting depends on the models used to describe flavor mixing in strangeonium.  It is also interesting to consider the prediction that one would obtain by assuming linear scaling ($d=2/3$).  In some sense this can be thought of as a non-relativistic, strong coupling limit.  In this case the predicted splitting becomes $M_{\Xi}(1 P)-M_{\Xi}(1 S)=$ 324.5 $\pm$ 4.6 MeV, in slightly better agreement with the experimental result.

\subsubsection{Symmetry Multiplets}

The angular momentum multiplets obtained in the RP model are listed in tables \ref{SmultipletsSpinless} and \ref{SmultipletsSpin} for the flavor-sextet states and tables \ref{ChicmultipletsOdd} and \ref{ChibmultipletsOdd} for the flavor-antitriplet states.  The masses for the $(\Xi_c,\Xi^\prime_c)$ and $(\Xi_b,\Xi^\prime_b)$ states presented here differ from those published in ref. \cite{RobertsPervin},  because the present spectrum has been recalculated using the $SU(3)_{heavy}$ flavor basis, but the original work used the familiar $SU(3)_{light}$ basis.  If (heavy) flavor-sextet and -antitriplet states are allowed to mix, the results would be identical with either choice of basis.  The results for strange diquarks in the spinless model show a remarkable similarity to the multiplet structure of $\Xi_{cc}$, $\Xi_{bc}$, and $\Xi_{bb}$ baryons.  The energy splittings due to pairwise mixing (section \ref{pairmix}) are slightly larger for baryons with strange diquarks than for diquarks comprised of $b$ and $c$ quarks.  This type of mixing causes the broken degeneracies of the $n_d^{2s_d+1}(l_\rho)$, $n_\lambda(l_\lambda)=1^{1}P$, $1P$ multiplets in table \ref{SmultipletsSpinless} and the $1^{3}P$, $1P$ multiplets in the ``Spinless'' column of tables \ref{ChicmultipletsOdd} and \ref{ChibmultipletsOdd}.  The largest pairwise splittings are in the $1^{1}P_1$, $1P_\frac{3}{2}$ multiplets.  For strange systems this splitting is 41 MeV (for the $\Xi$), and for doubly heavy systems the splitting is 24 MeV (for the $\Xi_{cc}$).  When spin interactions are included in the model, the largest spin splitting comes from the hyperfine splitting of the ground state multiplet.  As the mass of the heavy diquark system is increased from the $ss$, $cs$, $bs$, $cc$, $bc$, to the $bb$ system, this splitting decreases from 195, 115, 87, 74, 50, to 30 MeV, respectively.  In the limit where the $c$ and $s$ quarks are infinitely massive the statistics of the heavy diquark should be irrelevant.  Thus, the $\Xi_c^\prime$ in the ground state antitriplet should be degenerate with the $\Xi_c$ in the ground state sextet.  In the spinless RP model, the mass difference between these two states is 1 MeV. In the full model, these states have masses that are within 3\% of each other. 

Experimentally, only the ground state $\Xi_b$ has been established. For the $\Xi_c$ there are several excited states whose masses are known, but no quantum numbers have been assigned. It is therefore only possible to identify three of the angular momentum multiplets. The clearly identifiable $\Xi$ and $\Xi_c$ multiplets are listed in the tables discussed in the previous paragraph, beneath the corresponding model prediction.  For the $\Xi_c$ ground state multiplet, we have identified the states which are primarily composed of flavor-sextet heavy diquarks by comparing masses with predictions from the RP model.  Using baryon masses from experiment \cite{PDG}, the states belonging to the $\Xi$ ground state doublet are degenerate within 16\%, while states belonging to the first orbitally excited multiplet are within 8\% of each other.  The $\Xi_c$ ground state doublet is degenerate within 7\%.

At this point, it is possible to make a preliminary comparison between two different approaches for the $\Xi_c$ (or $\Xi_b$) baryon:  (1) treating the strange quark as a heavy quark and applying the heavy diquark symmetries that we have presented here; and (2) treating the strange quark as a light quark and applying HQS.  As discussed above, treating the strange quark as a heavy quark leads to the prediction that the three lowest lying $\Xi_{c}$/$\Xi^\prime_{c}$ states will be degenerate.  However, treating the strange quark as a light quark and applying HQS leads to the prediction that the three ground states will exist in a $J^p=(\hlf^+,\thlf^+)$ doublet with $J_l=1$ (here, $J_l$ refers to the spin of the light/strange degrees of freedom) and a $J^p=(\hlf^+)$ singlet with $J_l=0$.  The properties of these two states are unrelated because each multiplet carries light degrees of freedom with different total angular momentum.  To make this point more concrete, contrast the ground state of the $\Sigma_c$/$\Lambda_c$ with the ground state of the $\Xi_{c}$/$\Xi^\prime_{c}$.  The spin averaged mass of the $\Sigma_c$, $J^p=(\hlf^+,\thlf^+)$ HQS doublet is about 2500 MeV, the mass of the $\Lambda_c$, $J^p=(\hlf^+)$ HQS singlet is about 2290 MeV, and the difference is 210 MeV.  This difference in mass between the doublet (with $J_l=1$) and the singlet (with $J_l=0$) illustrates the idea that these two multiplets are not related by any underlying symmetry.  The analogous calculation for the $\Xi_{c}$/$\Xi^\prime_{c}$ baryons yields 150 MeV for the HQS doublet-singlet splitting suggesting the emergence of an additional symmetry.  This additional degeneracy which HQS does not predict is easily understood in terms of the decoupling of the total angular momentum of the heavy-strange diquark.

\begin{center}
\begin{table}[ht]
\caption{Symmetry multiplets for strange baryons in the RP model for flavor-sextet states with spin interactions removed.  The decoupling of the diquark total angular momentum predicts that the states within a multiplet will be degenerate.
\label{SmultipletsSpinless}}
\vspace{2mm}
\renewcommand{\arraystretch}{1.0}
\begin{tabular}{c c c c c }
\hline \toprule
Multiplet                         & $J^P$                       & \multicolumn{3}{c }{Mass (GeV)} \\ 
\hspace{1pt} $n_d^{2s_d+1}(l_\rho)_{J_d}$, $n_\lambda(l_\lambda)_{J_l}$ \hspace{1pt} 
                                        & $(|J_d-J_l|, ... ,J_d+J_l)$ & $\Xi     $ & $\Xi_{c}$ & $\Xi_{b}$\\ \hline
$1^{3}S_1$, $1S_\frac{1}{2}$             & $(\hlf^+,\thlf^+)$          & (1.451, 1.451) & (2.608, 2.608) & (5.959, 5.959) \\ 
$1^{1}P_1$, $1S_\frac{1}{2}$             & $(\hlf^-,\thlf^-)$          & (1.741, 1.741) & (2.865, 2.865) & (6.210, 6.210) \\ 
$2^{3}S_1$, $1S_\frac{1}{2}$             & $(\hlf^+,\thlf^+)$          & (1.968, 1.968) & (3.064, 3.064) & (6.404, 6.404) \\ 
$1^{3}D_1$, $1S_\frac{1}{2}$             & $(\hlf^+,\thlf^+)$          & (2.002, 2.002\footnotemark[1]) & (3.094, 3.094\footnotemark[1]) & (6.433, 6.433\footnotemark[1]) \\
$1^{3}D_2$, $1S_\frac{1}{2}$             & $(\thlf^+,\fhlf^+)$         & (2.002\footnotemark[1], 2.002) & (3.094\footnotemark[1], 3.095) & (6.433\footnotemark[1], 6.435) \\
$1^{3}D_3$, $1S_\frac{1}{2}$             & $(\fhlf^+,\shlf^+)$         & (2.002, 2.002) & (3.095, 3.094) & (6.435, 6.434) \\
$1^{3}S_1$, $1P_\frac{1}{2}$             & $(\hlf^-,\thlf^-)$          & (1.806, 1.806\footnotemark[2]) & (2.953, 2.953\footnotemark[2]) & (6.294, 6.294\footnotemark[2]) \\ 
$1^{3}S_1$, $1P_\frac{3}{2}$            & $(\hlf^-,\thlf^-,\fhlf^-)$   & (1.806, 1.806\footnotemark[2], 1.806) & (2.953, 2.953\footnotemark[2], 2.953) & (6.294, 6.294\footnotemark[2], 6.294) \\ 
$1^{1}P_1$, $1P_\frac{1}{2}$             & $(\hlf^+,\thlf^+)$          & (2.094\footnotemark[3], 2.078) & (3.207\footnotemark[3], 3.194) & (6.543\footnotemark[3], 6.531) \\ 
$1^{1}P_1$, $1P_\frac{3}{2}$            & $(\hlf^+,\thlf^+,\fhlf^+)$   & (2.053\footnotemark[3], 2.094, 2.077) & (3.174\footnotemark[3], 3.207, 3.192) & (6.512\footnotemark[3], 6.543, 6.529) \\
$1^{3}S_1$, $2S_\frac{1}{2}$          & $(\hlf^+,\thlf^+)$             & (2.091, 2.091) & (3.226, 3.226) & (6.560, 6.560) \\ 
$1^{3}S_1$, $1D_\frac{3}{2}$          & $(\hlf^+,\thlf^+,\fhlf^+)$     & (2.132, 2.132, 2.129\footnotemark[4]) & (3.268, 3.268, 3.262\footnotemark[4]) & (6.602,  6.602, 6.596\footnotemark[4]) \\ 
$1^{3}S_1$, $1D_\frac{5}{2}$          & $(\thlf^+,\fhlf^+,\shlf^+)$    & (2.132, 2.129\footnotemark[4], 2.132) & (3.268, 3.262\footnotemark[4], 3.268) & (6.602,  6.596\footnotemark[4], 6.602) \vspace{.5mm} \\  
\toprule
\footnotetext[1]{$^{\text{bcd}}$These states mix significantly.}
\end{tabular}
\end{table}
\end{center}
\begin{center}
\begin{table}[h]
\caption{Multiplets for strange baryons in the RP model for flavor-sextet states including spin dependent interactions.  When available, the experimental masses \cite{PDG} of these states are listed directly below the model predictions.  The decoupling of the diquark total angular momentum predicts that the states within a multiplet will be approximately degenerate.
\label{SmultipletsSpin}}
\vspace{2mm}
\renewcommand{\arraystretch}{1.0}
\begin{tabular}{c c c c c }
\hline \toprule
Multiplet                          & $J^P$                        & \multicolumn{3}{c }{Mass (GeV)} \\ 
\hspace{1pt} $n_d^{2s_d+1}(l_\rho)_{J_d}$, $n_\lambda(l_\lambda)_{J_l}$ \hspace{1pt} & $(|J_d-J_l|, ... ,J_d+J_l) $ & $\Xi$ & $\Xi_{c}$ & $\Xi_{b}$\\ \hline
$1^{3}S_1$, $1S_\frac{1}{2}$              & $(\hlf^+,\thlf^+)$           & (1.325, 1.520) & (2.537, 2.652) & (5.903, 5.990) \\[-5pt] 
 &         & (1.318, 1.533) & (2.469, 2.646)&-\\ 
$1^{1}P_1$, $1S_\frac{1}{2}$              & $(\hlf^-,\thlf^-)$           & (1.725, 1.759) & (2.861, 2.860) & (6.207, 6.206) \\[-5pt]
&        & (1.690, 1.823) & -& -\\
 $2^{3}S_1$, $1S_\frac{1}{2}$              & $(\hlf^+,\thlf^+)$           & (1.891, 2.021) & (3.030, 3.096) & (6.377, 6.424) \\ 
$1^{3}D_1$, $1S_\frac{1}{2}$              & $(\hlf^+,\thlf^+)$           & (2.055, 1.934\footnotemark[1]) & (3.119, 3.053\footnotemark[1]) & (6.450, 6.408\footnotemark[1]) \\
$1^{3}D_2$, $1S_\frac{1}{2}$              & $(\thlf^+,\fhlf^+)$          & (2.032\footnotemark[1], 2.025) & (3.113\footnotemark[1], 3.109) & (6.446\footnotemark[1], 6.443) \\
$1^{3}D_3$, $1S_\frac{1}{2}$              & $(\fhlf^+,\shlf^+)$          & (1.936, 2.035) & (3.053, 3.115) & (6.409, 6.447) \\
$1^{3}S_1$, $1P_\frac{1}{2}$              & $(\hlf^-,\thlf^-)$           & (1.826\footnotemark[3], 1.826\footnotemark[4]) & (2.976, 2.952\footnotemark[4]) & (6.311, 6.287\footnotemark[4]) \\ 
$1^{3}S_1$, $1P_\frac{3}{2}$             & $(\hlf^-,\thlf^-,\fhlf^-)$    & (1.811\footnotemark[3], 1.849\footnotemark[4], 1.883) & (2.952, 2.963\footnotemark[4], 2.970) & (6.287, 6.302\footnotemark[4], 6.307) \\ 
$1^{1}P_1$, $1P_\frac{1}{2}$              & $(\hlf^+,\thlf^+)$           & (2.108\footnotemark[2], 2.106) & (3.206\footnotemark[2], 3.195) & (6.523\footnotemark[2], 6.531) \\ 
$1^{1}P_1$, $1P_\frac{3}{2}$             & $(\hlf^+,\thlf^+,\fhlf^+)$   & (2.129\footnotemark[2]\footnotemark[5], 2.115, 2.137\footnotemark[6]) & (3.162\footnotemark[2]\footnotemark[5], 3.206, 3.195) & (6.508\footnotemark[2], 6.542, 6.530) \\
$1^{3}S_1$, $2S_\frac{1}{2}$          & $(\hlf^+,\thlf^+)$             & (2.014\footnotemark[5], 2.144) & (3.202\footnotemark[5], 3.256) & (6.538, 6.579) \\ 
$1^{3}S_1$, $1D_\frac{3}{2}$          & $(\hlf^+,\thlf^+,\fhlf^+)$     & (2.185, 2.149, 2.143\footnotemark[6]) & (3.277, 3.274, 3.267) & (6.609,  6.606, 6.596) \\ 
$1^{3}S_1$, $1D_\frac{5}{2}$          & $(\thlf^+,\fhlf^+,\shlf^+)$     & (2.142, 2.148\footnotemark[6], 2.148) & (3.268, 3.270, 3.275) & (6.598,  6.600, 6.607) \vspace{.5mm} \\ 
\toprule
\footnotetext[1]{$^{\text{bcdef}}$These states mix significantly.}
\end{tabular}
\end{table}
\end{center}
%
\begin{center}
\begin{table}[!h]
\caption{Symmetry multiplets for the RP model for $\Xi^\prime_{c}$ states.  When available, the experimental masses \cite{PDG} of these states are listed directly below the model predictions.  The decoupling of the diquark total angular momentum predicts that the states within a multiplet will be approximately degenerate.
\label{ChicmultipletsOdd}}
\vspace{2mm}
\renewcommand{\arraystretch}{1.0}
\begin{tabular}{c c c c}
\hline \toprule
Multiplet                   & $J^P$                        & \multicolumn{2}{c }{$\Xi^\prime_{c}$ Mass (GeV)} \\ 
\hspace{1pt} $n_d^{2s_d+1}(l_\rho)_{J_d}$, $n_\lambda(l_\lambda)_{J_l}$ \hspace{1pt} & $(|J_d-J_l|, ... ,J_d+J_l) $ & Spinless & Spin \\ \hline
$1^{1}S_0$, $1S_\frac{1}{2}$       & $(\hlf^+)$                   & (2.609) & (2.574)  \\[-5pt]
 &                 &  & (2.577)  \\ 

$1^{3}P_0$, $1S_\frac{1}{2}$       & $(\hlf^-)$                   & (2.865\footnotemark[1]) & (2.907\footnotemark[1])  \\ 
$1^{3}P_1$, $1S_\frac{1}{2}$       & $(\hlf^-,\thlf^-)$           & (2.865\footnotemark[1], 2.865) & (2.814\footnotemark[1], 2.899)  \\ 
$1^{3}P_2$, $1S_\frac{1}{2}$       & $(\thlf^-,\fhlf^-)$          & (2.865, 2.863) & (2.816, 2.890)  \\ 
$2^{1}S_0$, $1S_\frac{1}{2}$       & $(\hlf^+)$                   & (3.065) & (3.032)  \\ 
$1^{1}D_2$, $1S_\frac{1}{2}$       & $(\thlf^+,\fhlf^+)$          & (3.095, 3.095) & (3.091, 3.090)  \\
$1^{1}S_0$, $1P_\frac{1}{2}$       & $(\hlf^-)$                   & (2.953) & (2.927) \\ 
$1^{1}S_0$, $1P_\frac{3}{2}$       & $(\thlf^-)$                  & (2.953) & (2.927) \\ 
$1^{3}P_0$, $1P_\frac{1}{2}$       & $(\hlf^+)$                   & (3.192\footnotemark[2]) & (3.204\footnotemark[2]) \\ 
$1^{3}P_0$, $1P_\frac{3}{2}$       & $(\thlf^+)$                  & (3.205\footnotemark[3]) & (3.202\footnotemark[3]) \\ 
$1^{3}P_1$, $1P_\frac{1}{2}$       & $(\hlf^+,\thlf^+)$           & (3.205\footnotemark[2], 3.192\footnotemark[3]) 
                                                           & (3.216\footnotemark[2], 3.195\footnotemark[3])  \\ 
$1^{3}P_1$, $1P_\frac{3}{2}$       & $(\hlf^+,\thlf^+,\fhlf^+)$   & (3.205\footnotemark[2], 3.192\footnotemark[3], 3.192\footnotemark[4]) 
                                                           & (3.212, 3.178\footnotemark[3], 3.205\footnotemark[4])  \\ 
$1^{3}P_2$, $1P_\frac{1}{2}$       & $(\thlf^+,\fhlf^+)$          & (3.173\footnotemark[3], 3.192\footnotemark[4]) 
                                                           & (3.209, 3.177\footnotemark[4])  \\ 
$1^{3}P_2$, $1P_\frac{3}{2}$       & $(\hlf^+,\thlf^+,\fhlf^+,\shlf^+)$   & (3.173\footnotemark[2], 3.205\footnotemark[3], 3.205\footnotemark[4], 3.192) 
                                                                    & (3.116\footnotemark[2], 3.206\footnotemark[3], 3.199\footnotemark[4], 3.203)  \\ 
$1^{1}S_0$, $2S_\frac{1}{2}$       & $(\hlf^+)$                   & (3.221) & (3.219)  \\ 
$1^{1}S_0$, $1D_\frac{3}{2}$       & $(\thlf^+)$                  & (3.262) & (3.241)  \\ 
$1^{1}S_0$, $1D_\frac{5}{2}$       & $(\fhlf^+)$                  & (3.262) & (3.239)  \vspace{.5mm} \\ 
\toprule
\footnotetext[1]{$^{\text{bcd}}$These states mix significantly.}
\end{tabular}
\end{table}
\end{center}
\begin{center}
\begin{table}[!h]
\caption{Symmetry multiplets for the RP model for $\Xi^\prime_{b}$ states.  The decoupling of the diquark total angular momentum predicts that the states within a multiplet will be approximately degenerate.
\label{ChibmultipletsOdd}}
\vspace{2mm}
\renewcommand{\arraystretch}{1.0}
\begin{tabular}{c c c c}
\hline \toprule
Multiplet                   & $J^P$                        & \multicolumn{2}{c }{$\Xi^\prime_{b}$ Mass (GeV)} \\ 
\hspace{1pt} $n_d^{2s_d+1}(l_\rho)_{J_d}$, $n_\lambda(l_\lambda)_{J_l}$ \hspace{1pt} & $(|J_d-J_l|, ... ,J_d+J_l) $ & Spinless & Spin \\ \hline
$1^{1}S_0$, $1S_\frac{1}{2}$       & $(\hlf^+)$                   & (5.960) & (5.944)  \\ 
$1^{3}P_0$, $1S_\frac{1}{2}$       & $(\hlf^-)$                   & (6.210\footnotemark[1]) & (6.236\footnotemark[1])  \\ 
$1^{3}P_1$, $1S_\frac{1}{2}$       & $(\hlf^-,\thlf^-)$           & (6.210\footnotemark[1], 6.210) & (6.173\footnotemark[1], 6.228) \\ 
$1^{3}P_2$, $1S_\frac{1}{2}$       & $(\thlf^-,\fhlf^-)$          & (6.210, 6.209) & (6.173, 6.227)  \\ 
$2^{1}S_0$, $1S_\frac{1}{2}$       & $(\hlf^+)$                   & (6.406) & (6.390)  \\ 
$1^{1}D_2$, $1S_\frac{1}{2}$       & $(\thlf^+,\fhlf^+)$          & (6.435, 6.435) & (6.434, 6.432)  \\
$1^{1}S_0$, $1P_\frac{1}{2}$       & $(\hlf^-)$                   & (6.294) & (6.285) \\ 
$1^{1}S_0$, $1P_\frac{3}{2}$       & $(\thlf^-)$                  & (6.294) & (6.285) \\ 
$1^{3}P_0$, $1P_\frac{1}{2}$       & $(\hlf^+)$                   & (6.529\footnotemark[2]) & (6.540\footnotemark[2]) \\ 
$1^{3}P_0$, $1P_\frac{3}{2}$       & $(\thlf^+)$                  & (6.541\footnotemark[3]) & (6.539\footnotemark[3]) \\ 
$1^{3}P_1$, $1P_\frac{1}{2}$       & $(\hlf^+,\thlf^+)$           & (6.542\footnotemark[2], 6.529\footnotemark[3]) 
                                                           & (6.547\footnotemark[2], 6.531\footnotemark[3])  \\ 
$1^{3}P_1$, $1P_\frac{3}{2}$       & $(\hlf^+,\thlf^+,\fhlf^+)$   & (6.541\footnotemark[2], 6.529\footnotemark[3], 6.529\footnotemark[4]) 
                                                           & (6.545, 6.514\footnotemark[3], 6.536\footnotemark[4])  \\ 
$1^{3}P_2$, $1P_\frac{1}{2}$       & $(\thlf^+,\fhlf^+)$          & (6.510\footnotemark[3], 6.529\footnotemark[4]) 
                                                           & (6.544, 6.513\footnotemark[4])  \\ 
$1^{3}P_2$, $1P_\frac{3}{2}$       & $(\hlf^+,\thlf^+,\fhlf^+,\shlf^+)$   & (6.510\footnotemark[2], 6.540\footnotemark[3], 6.541\footnotemark[4], 6.529) 
                                                                    & (6.462\footnotemark[2], 6.542\footnotemark[3], 6.541\footnotemark[4], 6.539)  \\ 
$1^{1}S_0$, $2S_\frac{1}{2}$       & $(\hlf^+)$                   & (6.555) & (6.553)  \\ 
$1^{1}S_0$, $1D_\frac{3}{2}$       & $(\thlf^+)$                  & (6.596) & (6.587)  \\ 
$1^{1}S_0$, $1D_\frac{5}{2}$       & $(\fhlf^+)$                  & (6.596) & (6.586)  \vspace{.5mm} \\ 
\toprule
\footnotetext[1]{$^{\text{bcd}}$These states mix significantly.}
\end{tabular}
\end{table}
\end{center}

\subsubsection{Diquark Excitation and Flavor Symmetries}

The excitation symmetry emerges fairly clearly for diquarks comprising strange quarks. Table \ref{sFlavTable} shows the energy required to excite the light degrees of freedom in flavor-sextet DHBs in the spinless RP model for multiplets with two different diquark excitations: the ground state and the $n_d^{2s_d+1}(l_\rho)_{J_d}$=$1^1P_1$ multiplets. Table \ref{sFlavTableOdd} shows the analogous splittings for flavor-antitriplet states.  The excitation symmetry requires that the entries on the first row of table \ref{sFlavTable} should be the same as those on the second row, while those on the third row should be the same as those on the fourth row. In table \ref{sFlavTableOdd}, this symmetry means that entries on the first row should be repeated on the second, third and fourth rows, and entries on the fifth row should be repeated on the sixth, seventh and eighth rows. This symmetry should be broken by this particular model because of pair-wise interactions, but all of the excitation energies are within 17 MeV of each other for the flavor-sextet states, and within 25 MeV of each other for the antitriplet states.  The effects of pair-wise forces are therefore seen to be small, even for diquarks involving the strange quark. It is also interesting to note that these relations work better for states which have identical quarks in the heavy diquark.  This may be an indication that factorization emerges more readily for states containing such diquarks, and is a better approximation in those cases.

Tables \ref{sFlavTable} and \ref{sFlavTableOdd} also indicate the applicability of the heavy diquark flavor symmetry to strange diquarks.  The excitation energies should be equal for any diquark flavor.  Comparing the excitation energies of $\Xi$ baryons (table \ref{sFlavTable}) with those of $\Xi_{bb}$ baryons (table \ref{SuperFlavTablea}) reveals that the excitation energies are still remarkably similar.  The worst disagreements occur for the $\Xi_b$ baryons. Nevertheless, it is remarkable that these excitation energies are all so similar, as one would expect the $1/M_d$ corrections to these symmetry predictions to be appreciable for the $ss$ diquark.  One other trend that can be seen in table \ref{sFlavTable} is that the diquark flavor symmetry works better when the masses of the quarks comprising the diquark are closer to each other.  This trend is also apparent if we compare the splittings of $\Xi^\prime_{bc}$ baryons in table \ref{FlavTableOdd} with the splittings of $\Xi^\prime_{c}$ and $\Xi^\prime_{b}$ baryons in table \ref{sFlavTableOdd}.
\begin{center}

\begin{table}[h]
\caption{Energy required to excite the light degrees of freedom of a strange $SU(3)_{heavy}$ flavor-sextet baryon from the ground state, $J_l^{\pi_l}=\hlf^+$, to the excited state listed.  We have averaged over the masses in each multiplet.  This excitation energy should be independent of the spin, parity, and flavor of the heavy diquark, so that the energies in the first row should be the same as the corresponding ones in the second row, and the entries in the third row should equal those in the fourth row. These excitation energies should also be independent of the flavor of the quarks that comprise the strange diquark, so that all entries on a particular row should be equal. This is seen not only for the first four rows, but also for the last three rows.
\label{sFlavTable}}
\vspace{2mm}
\renewcommand{\arraystretch}{1.0}
\begin{tabular}{ccccccc}
\hline \toprule
Diquark state & &$\hphantom{aa}$ Upper Multiplet  & $\hphantom{a}$ Lower Multiplet $\hphantom{a}$ & \multicolumn{3}{c}{$m^*-m_0$ (MeV)}\\\cline{5-7}
\hspace{1pt}$n_d^{2s_d+1}(l_\rho)_{J_d}$& $\hphantom{ae}n_\lambda(l_\lambda)_{J_l}\hphantom{ab}$& $\hphantom{\hlf^-,}$($m^*$)   & ($m_0$) \hspace{1pt} & $\hphantom{a}\Xi\hphantom{a}$ & $\hphantom{a}\Xi_{c}\hphantom{a}$ & $\hphantom{a}\Xi_{b}\hphantom{a}$ \\ \hline 

$1^3S_1$& $1P_\hlf$         & $\hphantom{\hlf^-,}(\hlf^-,\thlf^-)$& $(\hlf^+,\thlf^+)$    & 355 & 345 & 335 \\ 
$1^1P_1$& $1P_\hlf$         &$\hphantom{\hlf^-,}(\hlf^+,\thlf^+)$ & $(\hlf^-,\thlf^-)$    & 345 & 336 & 327 \\ 
$1^3S_1$& $1P_\thlf$        & $(\hlf^-,\thlf^-,\fhlf^-)$& $(\hlf^+,\thlf^+)$ & 355 & 345 & 335 \\ 
$1^1P_1$& $1P_\thlf$        &$(\hlf^+,\thlf^+,\fhlf^+)$& $(\hlf^-,\thlf^-)$   & 334 & 326 & 318  \\ \\
$1^3S_1$& $1D_\thlf$        & $(\hlf^+,\thlf^+,\fhlf^+)$& $(\hlf^+,\thlf^+)$    & 680 & 658 & 641 \\
$1^3S_1$& $1D_\fhlf$        &$(\thlf^+,\fhlf^+,\shlf^+)$&$(\hlf^+,\thlf^+)$     & 680 & 658 & 641 \\
$1^3S_1$& $2S_\hlf$         & $\hphantom{\hlf^-,}(\hlf^+,\thlf^+)$ & $(\hlf^+,\thlf^+)$    & 640 & 658 & 601 \\

\toprule

\end{tabular}
\end{table}
\end{center}
\begin{center}
\begin{table}[!h]
\caption{Energy required to excite the light degrees of freedom of a strange $SU(3)_{heavy}$ flavor-antitriplet baryon from the ground state, $J_l^{\pi_l}=\hlf^+$, to the excited state listed.  We have averaged over the masses in each multiplet.  This excitation energy should be independent of the spin, parity, and flavor of the heavy diquark, so that the energies in the rows 1, 2, 3, and 4 should be equal, and the entries in rows 5, 6, 7, and 8. These excitation energies should also be independent of the flavor of the quarks that comprise the strange diquark, so that all entries on a particular row should be equal. This is seen not only for the first eight rows, but also for the last three rows.
\label{sFlavTableOdd}}
\vspace{2mm}
\renewcommand{\arraystretch}{1.0}
\begin{tabular}{cccccc}
\hline \toprule
Diquark state & & Upper Multiplet  & $\hphantom{a}$  Lower Multiplet $\hphantom{a}$ & \multicolumn{2}{c}{$m^*-m_0$ (MeV)}\\\cline{5-6}
\hspace{1pt}$n_d^{2s_d+1}(l_\rho)_{J_d}$& $\hphantom{ae}n_\lambda(l_\lambda)_{J_l}\hphantom{ab}$& ($m^*$)   \hspace{1pt} & ($m_0$) \hspace{1pt} & $\hphantom{a}\Xi_{c}\hphantom{a}$ & $\hphantom{a}\Xi_{b}\hphantom{a}$ \\ \hline 

$1^1S_0$& $1P_\hlf$             &$\hphantom{\hlf^-,\hlf^-,\hlf^+,}(\hlf^-)$ &$\hphantom{\hlf^-,\hlf^-,}(\hlf^+)$ & 344 & 334  \\ 
$1^3P_0$& $1P_\hlf$             &$\hphantom{\hlf^-,\hlf^-,\hlf^+,}(\hlf^+)$  & $\hphantom{\hlf^-,\hlf^-,}(\hlf^-)$ & 327 & 319  \\ 
$1^3P_1$& $1P_\hlf$             &$\hphantom{\hlf^-,\hlf^-,}(\hlf^+,\thlf^+)$& $\hphantom{\hlf^-,}(\hlf^-,\thlf^-)$ & 334 & 326  \\ 
$1^3P_2$& $1P_\hlf$             &$\hphantom{\hlf^-,\hlf^-,}(\thlf^+,\fhlf^+)$& $(\hlf^-,\thlf^-,\fhlf^-)$ & 319 & 310  \\ 
$1^1S_0$& $1P_\thlf$            & $\hphantom{\hlf^-,\hlf^-,\hlf^+,}(\thlf^-)$& $\hphantom{\hlf^-,\hlf^-,}(\hlf^+)$ & 344 & 334  \\ 
$1^3P_0$& $1P_\thlf$            &$\hphantom{\hlf^-,\hlf^-,\hlf^+,}(\thlf^+)$& $\hphantom{\hlf^-,\hlf^-,}(\hlf^-)$  & 340 & 331   \\ 
$1^3P_1$& $1P_\thlf$            & $\hphantom{\hlf^-,}(\hlf^+,\thlf^+,\fhlf^+)$& $\hphantom{\hlf^-,}(\hlf^-,\thlf^-)$ & 331 & 323  \\ 
$1^3P_2$& $1P_\thlf$            &$(\hlf^+,\thlf^+,\fhlf^+,\shlf^+)$ & $(\hlf^-,\thlf^-,\fhlf^-)$ & 330 & 320   \\ \\
$1^1S_0$& $1D_\thlf$            &$\hphantom{\hlf^-,\hlf^-,\hlf^+,}(\thlf^+)$&  $\hphantom{\hlf^-,\hlf^-,}(\hlf^+)$ & 653 & 636  \\
$1^1S_0$& $1D_\fhlf$            & $\hphantom{\hlf^-,\hlf^,\hlf^+-,}(\fhlf^+)$ &$\hphantom{\hlf^-,\hlf^-,}(\hlf^+)$  & 653 & 636  \\
$1^1S_0$& $2S_\hlf$             &$\hphantom{\hlf^-,\hlf^-,\hlf^+,}(\hlf^+)$& $\hphantom{\hlf^-,\hlf^-,}(\hlf^+)$ & 612 & 595  \\ 
\toprule
\end{tabular}
\end{table}
\end{center}

\subsubsection{Superflavor Symmetry}
\begin{center}
\begin{figure}
\caption{Energy required to excite the light degrees of freedom of singly heavy mesons from the ground state, $J_l^{\pi_l}=\hlf^+$, to the $J_l^{\pi_l}=\thlf^-$ state from experiment \cite{PDG}.  This is plotted against the inverse mass of the heavy quark using quark masses from the RP model \cite{RobertsPervin}.
\label{HQSplot}}
\includegraphics{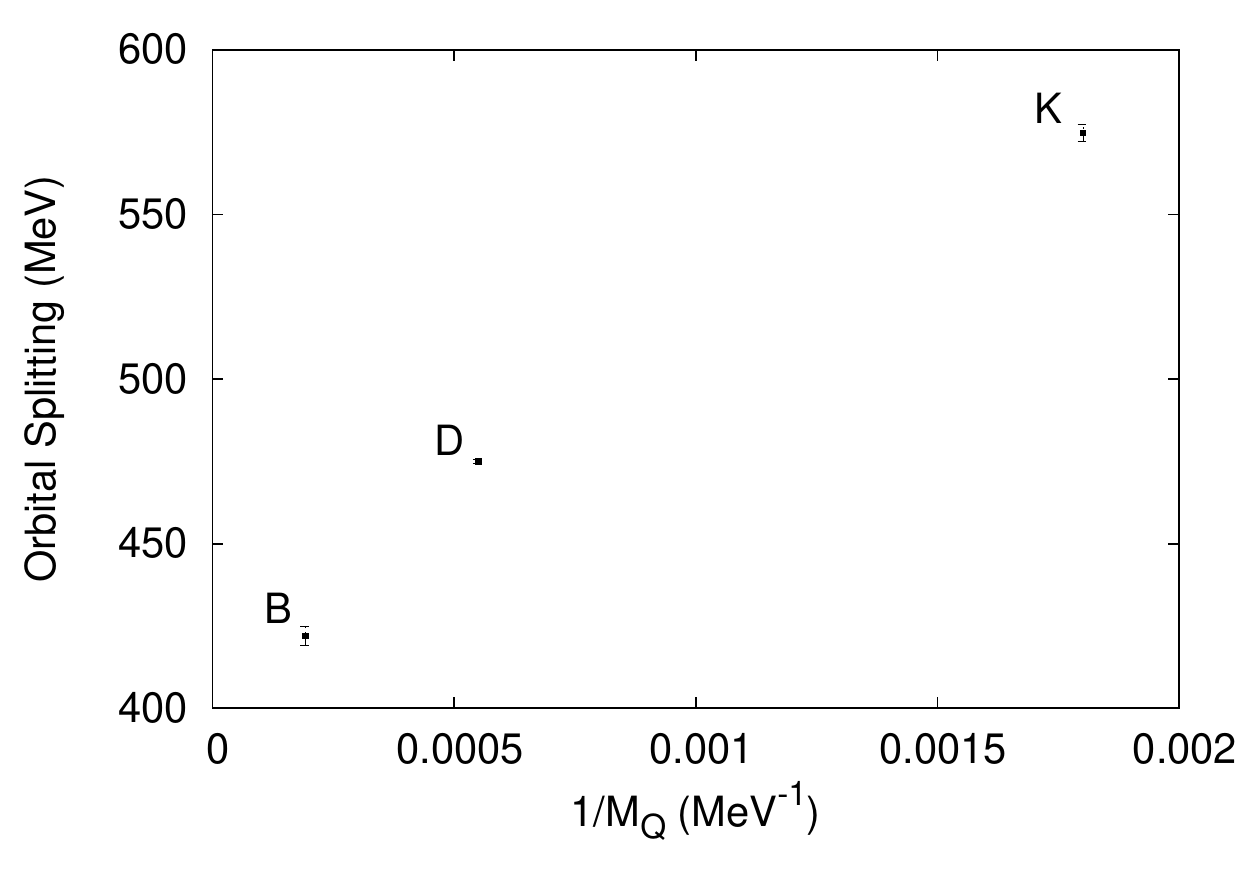}
\end{figure}
\end{center}

In section \ref{SupFlavResults} we used the superflavor symmetry to estimate that the energy required to excite the light degrees of freedom of a DHB from the ground state to the $J_l^{\pi_l}=\thlf^-$ excited state is about 400 MeV.  However, this estimate is only valid for very heavy diquarks.  For strange diquarks, one can estimate the diquark mass as $M_d \sim M_{Q_1}+M_{Q_2}$.  In the RP model this gives a mass of about 1.1 GeV for the $ss$ diquark.  As such, corrections due to its finite mass could be quite important.  This is illustrated in fig. \ref{HQSplot} which shows the energy of orbital excitations for singly heavy mesons from experiment plotted against the inverse heavy quark masses taken from the RP model.  These three data points come remarkably close to lying on a straight line considering the relatively small mass of the strange quark. These orbital splittings may be parametrized up to second order in the heavy quark expansion as $\Delta M=K_2/M^2_Q+K_1/M_Q+\Delta M_0$, where $K_1$ and $K_2$ are constants which are chosen to reflect experimental results and $\Delta M_0$ is the value of the orbital splitting that emerges in the infinitely massive heavy quark limit.  Fitting the heavy meson data from table \ref{FlavTableThlfm} gives, $K_2=-4.26 \times 10^7$ MeV$^{3}$, $K_1=1.80 \times 10^5$ MeV$^{2}$, and $\Delta M_0=388.9$ MeV.  HQET arguments indicate that $K_1$ should be on the order of $\Lambda^2_{QCD} \sim 10^5$ MeV$^{2}$ and $K_2$ should be on the order of $\Lambda^3_{QCD} \sim 10^7$ MeV$^{3}$.  Thus the values of the correction coefficients which we have obtained in this phenomenological approach seem to indicate that inclusion of corrections up to ${\cal O}(1/M_Q^2)$ would describe these systems quite well.  Because these corrections cannot be neglected in singly heavy meson systems one may speculate that they are also important for systems with strange diquarks. A systematic treatment of these corrections is left for possible future work. 

\section{Conclusions}

We have examined the systematics and symmetries of the spectra of baryons containing two heavy quarks, particularly as they pertain to the model of Roberts and Pervin \cite{RobertsPervin}. We have argued for an extension of the superflavor symmetry, which was proposed by Savage and Wise \cite{SavageSpectrum}, to include excitations of the heavy diquark, and we have examined some of the consequences of this extension for the spectra of DHBs. In this extension, the total angular momentum of the heavy diquark decouples from the light degrees of freedom independent of the excitation state of the heavy diquark.  Using the Feynman-Hellman Theorem, we have derived an equation which describes how the excitation energy of a doubly heavy meson or a heavy diquark scales with the masses of the heavy quarks and the strength of the interaction.  The spectra of bottomonium and charmonium indicate that the excitation energy of these heavy-heavy systems are independent of the heavy quark mass, and grow linearly with the interaction strength.  We have proposed a relation between excitations of heavy diquarks and those of doubly heavy mesons.  This relation depends on the form of the confining forces.  When combined with factorization, this leads to a relation between the spectra of DHBs and the spectra of doubly heavy mesons. Even though the model of ref. \cite{RobertsPervin} was not constructed using any of the symmetries we discussed, we found that these symmetries nevertheless emerged. 

We have also explored the applicability of these symmetries to systems with strange quarks.  Results from the RP model for strange diquarks are remarkably similar to results for DHBs. There is some indication that the consequences of factorization do not emerge as well for $\Xi_{c}$ and $\Xi_{b}$ baryons as they do for DHBs and the $\Xi$.  The fact that {\it any} of the consequences of factorization emerge is rather surprising and suggests that (at least for this particular model) a point-like diquark is not necessary for the system to factorize.  $\Xi_{c}$ and $\Xi_{b}$ baryons are also interesting, in that two different descriptions may be possible. On one hand, one may interpret their spectra in terms of the superflavor and supercolor symmetries that we have presented by considering these systems to be composed of a `heavy' diquark.  On the other hand, one may think of the strange quark as a light quark, in which case, the spectra can be interpreted in terms of the HQS and the $SU(3)_{light}$ flavor symmetry.  Whether these very different descriptions are complimentary or contradictory remains to be determined.  Our estimate of the lowest orbital splitting of the $\Xi$ using the supercolor symmetry agreed with the experimental data.  If this is not coincidence due to our choices of phenomenological models for strangeonium mixing, then it provides some evidence for the existence of this symmetry. Additionally, the orbital splittings of singly heavy mesons seem to show that HQS is applicable (phenomenologically at least) to strange quark systems.  By extension, we expect that the superflavor symmetry will provide reasonable predictions for systems containing strange diquarks as well as heavy diquarks, but this remains to be tested.

This work has concocted a description of DHBs in which the entire spectrum may be approximately constructed from the spectra of singly and doubly heavy mesons once the mass of the ground state DHB has been measured.  We have neglected spin-dependent interactions and assumed that factorization is valid.  We have also ignored the mixing of heavy diquark flavor-singlets and diquark flavor-triplets due to hyperfine interactions.  For the ratios involving doubly heavy mesons, we have also assumed that it is reasonable to use a non-relativistic potential model, or rather that such a model exists which captures the essence of QCD interactions for these systems.  It would be interesting if ratios similar to the ones we have derived could be extracted from NRQCD.  We have also assumed that the strength of interactions for a doubly heavy meson is exactly twice that of a heavy diquark. However, to our knowledge, there is no rigorous demonstration that this holds in general.  It would also be interesting to see if there is a firmer conceptual foundation for the symmetry between doubly heavy mesons and heavy diquarks.  Simplistically, this is a symmetry between a heavy quark and a heavy antiquark in a color singlet and two heavy quarks in a color antitriplet, but we have ignored how this relates to charge conjugation and parity inversion.  The simple result that for $bb$, $cc$, and $cs$ systems the energy scaling index $d \approx 1$ reveals an important fact regarding doubly heavy hadron spectroscopy that models for baryons should not ignore -- the energy required to orbitally excite two heavy quarks is nearly independent of the heavy quark masses, at least, for the range of masses that encompass these diquarks.  Moreover, the heavy quarks available to hadron physicists are not massive enough for the spectrum to behave in a Coulombic manner.  We hope that this work has demonstrated the insights into QCD which the study of doubly heavy hadrons offers. We propose to examine other aspects of the phenomenology of heavy hadrons in the framework of the symmetries and systematics that we have discussed.

\section*{Acknowledgment} We gratefully acknowledge the support of the Department of Physics, the College of Arts and Sciences, and the Office of Research at Florida State University.
This research is supported by the U.S. Department of Energy under contract
DE-SC0002615.

\newif\ifmultiplepapers
\def\beginpapers{\multiplepaperstrue}
\def\endpapers{\multiplepapersfalse}  
\def\journal#1&#2(#3)#4{\rm #1~{\bf #2}\unskip, \rm  #4 (19#3)}
\def\trjrnl#1&#2(#3)#4{\rm #1~{\bf #2}\unskip, \rm #4 (19#3)}
\def\baps{\journal {Bull.} {Am.} {Phys.} {Soc.}&}
\def\jap{\journal J. {Appl.} {Phys.}&}
\def\prl{\journal {Phys.} {Rev.} {Lett.}&}
\def\pl{\journal {Phys.} {Lett.}&}
\def\pr{\journal {Phys.} {Rev.}&}
\def\np{\journal {Nucl.} {Phys.}&}
\def\rmp{\journal {Rev.} {Mod.} {Phys.}&}
\def\jmp{\journal J. {Math.} {Phys.}&}
\def\rmm{\journal {Revs.} {Mod.} {Math.}&}
\def\jetp{\journal {J.} {Exp.} {Theor.} {Phys.}&}
\def\sjetp{\trjrnl {Sov.} {Phys.} {JETP}&}
\def\dokl{\journal {Dokl.} {Akad.} Nauk USSR&}
\def\spd{\trjrnl {Sov.} {Phys.} {Dokl.}&}
\def\tmf{\journal {Theor.} {Mat.} {Fiz.}&}
\def\snp{\trjrnl {Sov.} J. {Nucl.} {Phys.}&}
\def\hpa{\journal {Helv.} {Phys.} Acta&}
\def\yf{\journal {Yad.} {Fiz.}&}
\def\zp{\journal Z. {Phys.}&}
\def\anp{\journal {Adv.} {Nucl.} {Phys.}&}
\def\ap{\journal {Ann.} {Phys.}&}
\def\am{\journal {Ann.} {Math.}&}
\def\nc{\journal {Nuo.} {Cim.}&}
\def\etal{{\sl et al.}}
\def\pre{\journal {Phys.} {Rep.}&}
\def\pca{\journal Physica (Utrecht)&}
\def\prs{\journal {Proc.} R. {Soc.} London &}
\def\jcp{\journal J. {Comp.} {Phys.}&}
\def\pna{\journal {Proc.} {Nat.} {Acad.}&}
\def\jpg{\journal J. {Phys.} G (Nuclear Physics)&}
\def\fort{\journal {Fortsch.} {Phys.}&}
\def\jfa{\journal {J.} {Func.} {Anal.}&}
\def\cmp{\journal {Comm.} {Math.} {Phys.}&}


\newpage

\begin{thebibliography}{0}
\normalsize
\baselineskip = 0.2 in
\parskip 0pt

\bibitem{SavageSpectrum} M.~J.~Savage and M.~B.~Wise, Phys.\ Lett.\  B {\bf 248}, 177 (1990).

\bibitem{Cohen:2006jg}
  T.~D.~Cohen and P.~M.~Hohler,
  Phys.\ Rev.\  D {\bf 74}, 094003 (2006).

\bibitem{RobertsPervin} W. Roberts and M. Pervin, Int. \ J. \ Mod. \ Phys.\ A {\bf 23}, 2817 (2008), arXiv:0711.2492.

\bibitem{Fleck:1989mb}
  S.~Fleck and J.~M.~Richard,
  Prog.\ Theor.\ Phys.\  {\bf 82}, 760 (1989).

\bibitem{Stong} M. L. Stong, arXiv:9505217 [hep-ph].

\bibitem{Gershtein:1998sx}
  S.~S.~Gershtein, V.~V.~Kiselev, A.~K.~Likhoded and A.~I.~Onishchenko,
  Mod.\ Phys.\ Lett.\  A {\bf 14}, 135 (1999).

\bibitem{gershtein00} S.S. Gershtein, V.V. Kiselev, A.K. Likhoded, and 
A.I. Onishchenko, Phys. Rev. D {\bf 62}, 054021 (2000).

\bibitem{Kiselev:Omegas}
V. V. Kiselev, A. K. Likhoded, O. N. Pakhomova, V. A. Saleev, Phys. Rev. D {\bf 66}, 
     034030 (2002), arXiv:0206140. 

\bibitem{Ebert:2002ig}
  D.~Ebert, R.~N.~Faustov, V.~O.~Galkin and A.~P.~Martynenko,
  Phys.\ Rev.\  D {\bf 66}, 014008 (2002).

\bibitem{Majethiya} A.~Majethiya, B.~Patel, A.~K.~Rai, and P.~C.~Vinodkumar,  arXiv:0809.4910 [hep-ph].

\bibitem{Giannuzzi} F. Giannuzzi, Phys. Rev. D {\bf 79}, 094002 (2009).


\bibitem{Rujula75} A. De Rujula, H. Georgi and S. L. Glashow, Phys. Rev. D {\bf 12}, 147 (1975).


\bibitem{Bagan:1994dy}
  E.~Bagan, H.~G.~Dosch, P.~Gosdzinsky, S.~Narison and J.~M.~Richard,
  Z.\ Phys.\  C {\bf 64}, 57 (1994).
               

\bibitem{SilvestreBrac:1996wp}
  B.~Silvestre-Brac,
  Prog.\ Part.\ Nucl.\ Phys.\  {\bf 36}, 263 (1996).

\bibitem{Richard:1996za}
  J.~M.~Richard,
  Nucl.\ Phys.\ Proc.\ Suppl.\  {\bf 50}, 147 (1996).

\bibitem{itoh00} C. Itoh, T. Minamikawa, K. Miura, and T. Watanabe,
Phys. Rev. D {\bf 61}, 057502 (2000).

\bibitem{vijande04} J. Vijande, H. Garcilazo, A. Valcarce, and F. Fern\'andez,
Phys. Rev. D {\bf 70}, 054022 (2004).

\bibitem{Albertus:2006ya}
  C.~Albertus, E.~Hernandez, J.~Nieves and J.~M.~Verde-Velasco,
  Eur.\ Phys.\ J.\  A {\bf 32}, 183 (2007).

\bibitem{Zheng10} W. Zheng, H.R. Pang,  Mod. Phys. Lett. A {\bf 25}, 2077 (2010).



\bibitem{martynenko}
  A.~P.~Martynenko,
  arXiv:0708.2033 [hep-ph].

\bibitem{Patel08}
B. Patel,  A. K. Rai, P.C. Vinodkumar,  . Feb 2008. 10pp. 
Pramana {\bf 70}, 797 (2008),  arXiv:0802.4408 [hep-ph].

\bibitem{narodetskii02} I.M. Narodetskii and M.A. Trusov, 
Phys. At. Nucl. {\bf 65}, 917 (2002)  (Yad. Fiz. {\bf 65}, 944 (2002)).

\bibitem{Narodetskii:2002ks}
  I.~M.~Narodetskii, A.~N.~Plekhanov and A.~I.~Veselov,
  JETP Lett.\  {\bf 77}, 58 (2003)
  [Pisma Zh.\ Eksp.\ Teor.\ Fiz.\  {\bf 77}, 64 (2003)].




\bibitem{Likhoded} A. K. Likhoded, Phys. of At. and Nuc. {\bf 72}, 529 (2009).

\bibitem{Ebert:1996ec}
  D.~Ebert, R.~N.~Faustov, V.~O.~Galkin, A.~P.~Martynenko and V.~A.~Saleev,
  Z.\ Phys.\  C {\bf 76}, 111 (1997).

\bibitem{Ebert:2005ip}
  D.~Ebert, R.~N.~Faustov, V.~O.~Galkin and A.~P.~Martynenko,
  Phys.\ Atom.\ Nucl.\  {\bf 68}, 784 (2005)
  [Yad.\ Fiz.\  {\bf 68}, 817 (2005)].

\bibitem{Gerasyuta:1999pc}
  S.~M.~Gerasyuta and D.~V.~Ivanov,
  Nuovo Cim.\  A {\bf 112}, 261 (1999).


\bibitem{Lyubovitskij:2003pn}
  V.~E.~Lyubovitskij, A.~Faessler, T.~Gutsche, M.~A.~Ivanov and J.~G.~Korner,
  Prog.\ Part.\ Nucl.\ Phys.\  {\bf 50}, 329 (2003).

\bibitem{faessler}
A.~Faessler, T.~Gutsche, M.~A.~Ivanov, J.~G.~Korner, V.~E.~Lyubovitskij, D.~Nicmorus and K.~Pumsa-ard,
  Phys.\ Rev.\  D {\bf 73}, 094013 (2006).

\bibitem{Gerasyuta:2008zy}
  S.~M.~Gerasyuta and E.~E.~Matskevich,
  arXiv:0803.3497 [hep-ph].


\bibitem{He2004}
D. H. He, K. Qian, Y. B. Ding, X. Q. Li and P. N. Shen, Phys. Rev. D {\bf 70}, 094004 
     (2004) arXiv:0403301 [hep-ph]. 

\bibitem{tong2000} S. P. Tong, Y. B. Ding, X. H. Guo, H. Y. Jin, X. Q. Li, P. N. Shen and R. Zhang, 
     Phys. Rev. D {\bf 62}, 054024 (2000), arXiv:9910259 [hep-ph]. 


\bibitem{Weng2011}
M.-H. Weng, X.-H. Guo, A.W. Thomas,  
Phys. Rev. D {\bf 83}, 056006 (2011), arXiv:1012.0082 [hep-ph]. 

\bibitem{Bagan:1992za}
  E.~Bagan, M.~Chabab and S.~Narison,
  Phys.\ Lett.\  B {\bf 306}, 350 (1993).


\bibitem{Kisilev2001} V. V. Kiselev and A. E. Kovalsky,
Phys Rev. D {\bf 64}, 014002 (2001).

\bibitem{Zhang:2008} J. R. Zhang and M. Q. Huang, Phys. Rev. D {\bf 78}, 094007 (2008), arXiv:0810.5396 [hep-ph].

\bibitem{Alb:Nuc} R. M. Albuquerque and S. Narison, Nucl. Phys. (Proc. Suppl.) 207-208, 265 (2010), 
    arXiv:1009.2428 [hep-ph]. 

\bibitem{Alb:PhLett} R. M. Albuquerque and  S. Narison, Phys. Lett. B {\bf 694}, 217 (2010), arXiv:1006.2091 [hep-ph]. 

\bibitem{tang:2011}  L. Tang, X. Yuan, C. Qiao, X. Li, arXiv:1104.4934 [hep-ph].



\bibitem{Brambilla05}
  N.~Brambilla, A.~Vairo and T.~R\"{o}sch,
  Phys.\ Rev.\  D {\bf 72}, 034021 (2005).

 \bibitem{Hu2005}
  J. Hu and T. Mehen,
  Phys.\ Rev.\  D {\bf 73}, 054003 (2006).

 \bibitem{Fleming2006}
  S. Fleming and T. Mehen,
  Phys.\ Rev.\  D {\bf 73}, 034502 (2006).

 \bibitem{Mehen2006}
  T. Mehen and B. C. Tiburzi,
  Phys.\ Rev.\  D {\bf 74}, 054505 (2006).

 
\bibitem{lewis01} R. Lewis, N. Mathur, and R. M. Woloshyn, 
Phys. Rev D {\bf 64}, 094509 (2001).
  
\bibitem{Flynn:2003vz}
  J.~M.~Flynn, F.~Mescia and A.~S.~B.~Tariq  [UKQCD Collaboration],
  JHEP {\bf 0307}, 066 (2003).

\bibitem{Na:2007pv}
  H.~Na and S.~A.~Gottlieb,
  arXiv:0710.1422 [hep-lat].


\bibitem{Mathur:2001id}
  N.~Mathur, R.~Lewis and R.~M.~Woloshyn,
  Nucl.\ Phys.\ Proc.\ Suppl.\  {\bf 106}, 400 (2002).

\bibitem{mathur02} N. Mathur, R. Lewis, and R. M. Woloshyn, 
Phys. Rev. D {\bf 66}, 014502 (2002).


\bibitem{Lichtenberg:1995kg}
  D.~B.~Lichtenberg, R.~Roncaglia and E.~Predazzi,
  Phys.\ Rev.\  D {\bf 53}, 6678 (1996).

 \bibitem{Roncaglia:1995az}
  R.~Roncaglia, D.~B.~Lichtenberg and E.~Predazzi,
  Phys.\ Rev.\  D {\bf 52}, 1722 (1995).


\bibitem{kiselev02} V.V. Kiselev and A.K. Likhoded, Phys. Usp. {\bf 45}, 455 (2002)
(Usp. Fiz. Nauk {\bf 172}, 497 (2002)). arXiv:0103169.

\bibitem{Castro}  A. S. de Castro and M. F. Sugaya, Eur. J. Phys. {\bf 14}, 259 (1993).

\bibitem{Nakamura} A.~Nakamura and T.~Saito, Phys.\ Lett.\  B {\bf 621}, 171 (2005).

\bibitem{Quigg}  C. Quigg and J. L. Rosner, Phys. Rep. {\bf 56}, 167 (1979).

\bibitem{Cohen1979} I. Cohen, and H. Lipkin, Phys. Lett. B {\bf 84}, 323 (1979).

\bibitem{Kwong} W. Kwong and J. L. Rosner, Phys.\ Rev.\  D {\bf 44}, 212 (1991).

\bibitem{Lipkin1993} H. Lipkin, Phys. Lett. B {\bf 319}, 276 (1993).

\bibitem{Roncaglia} R. Roncaglia, A. Dzierba, D. B. Lichtenberg, and E. Predazzi, Phys. Rev. D {\bf 51}, 1248 (1995).


\bibitem{IsgurWise} N.~Isgur and M.~B.~Wise, Phys.\ Rev.\ Lett.\ {\bf 66}, 1130 (1991).

\bibitem{GeorgiFlav} H.~Georgi and M.~B.~Wise, Phys.\ Lett.\  B {\bf 243}, 279 (1990).

\bibitem{Thacker91} B. A. Thacker, G. P. Lepage, Phys. Rev. D {\bf 43}, 196 (1991).


\bibitem{PDG}  K. Nakamura et al. (Particle Data Group), J. Phys. G {\bf 37}, 075021 (2010).

\bibitem{Belle} I. Adachi et al. (Belle Collaboration), arXiv:1103.3419 [hep-ex].

\bibitem{BABAR} B. Aubert et al. (BABAR Collaboration), PRL {\bf 101}, 071801 (2008).

\bibitem{EbertMesons} D.~Ebert, R.~N.~Faustov and V.~O.~Galkin, Phys.\ Rev.\  D {\bf 67}, 014027 (2003). arXiv:0210381 [hep-ph].

\bibitem{BrambillaRev}
  N.~Brambilla, A. ~Pineda, J. ~Soto and A.~Vairo,
   Rev.\ Mod.\ Phys.\  {\bf 77}, 1423 (2005).

\bibitem{Thomas}  C. E. Thomas, J. High Energy Phys. {\bf 10}, 026 (2007).

\bibitem{CloseZhao} F. E. Close and Q. Zhao, Phys. Rev. D {\bf 71}, 094022 (2005).

\bibitem{CloseKirk} F. E. Close and A. Kirk, Z. Phys. C {\bf 76}, 469 (1997).

\bibitem{Dudek} J. J. Dudek, et al., Phys. Rev. D {\bf 83}, 111502 (2011).

  \end{thebibliography}
\end{document}